\documentclass[11pt,a4paper]{article}
\pdfoutput=1
\usepackage{jheppub}
\usepackage{amsmath}
\usepackage{verbatim}
\usepackage{graphicx}
\usepackage{mathrsfs}
\usepackage{appendix}
\usepackage{caption}
\usepackage{float}
\usepackage{subfig}
\usepackage{mathtools}
\usepackage{jheppub}
\usepackage{amsfonts}

\usepackage{amssymb}
\usepackage{inputenc, array}
\usepackage{textcomp}
\usepackage[dvipsnames]{xcolor}

\newcommand{\z}{{\bar z}}
\newcommand{\h}{{\bar h}}

\renewcommand\O{{\mathcal{O}}}
\newcommand{\be}[1]{ \begin{equation}\label{#1} }
\newcommand{\ee}{\end{equation}}
\newcommand{\bea}[1]{\begin{eqnarray}\label{#1} }
\newcommand{\eea}{\end{eqnarray}}
\newcommand{\bes}{\begin{subequations}}
\newcommand{\ees}{\end{subequations}}

\newcommand{\p}{\partial}

\DeclarePairedDelimiterX\braket[2]{\langle}{\rangle}{#1 \delimsize\vert #2}

\definecolor{mycolor}{RGB}{24, 116, 148}
\hypersetup{colorlinks=true,citecolor=mycolor,linkcolor=mycolor, linktocpage,urlcolor=mycolor}

\newcommand{\SL}{\mbox{SL}(2,\mathbb{R})}

\title{Anisotropic conformal Carroll field theories and their gravity duals}

\author{Emilie Despontin,} \author{Stéphane Detournay,} \author{Sudipta Dutta,} \author{and Dima Fontaine. \\}

\affiliation{Physique Théorique et Mathématique and International Solvay Institutes,\\
Université Libre de Bruxelles (ULB), C.P. 231, 1050 Brussels, Belgium.}
\emailAdd{emilie.despontin@ulb.be, stephane.detournay@ulb.be,sudipta.dutta@ulb.be, dima.fontaine@ulb.be}

\abstract{
We investigate anisotropic conformal Carroll field theories and their holographic duals. On the field theory side, we focus on the case with scaling exponent $z=0$ in two and three spacetime dimensions. These theories exhibit infinite-dimensional symmetry algebras, including supertranslations and superrotations, and are closely related to, but distinct from, Warped Conformal Field Theories. We construct the associated Carrollian stress tensor, derive its transformation properties, and analyse the structure of correlation functions under different choices of vacua. On the gravity side, we identify three and four-dimensional plane wave geometries whose isometry algebras realise the two- and three-dimensional Carroll algebra and anisotropic scale transformations. We propose, for each scaling exponent, a phase space of asymptotically-plane wave spacetimes and show that the residual diffeomorphisms reproduce the expected conformal Carroll field theory algebra, establishing a framework for anisotropic Carrollian holography.

}

\preprint{}

\begin{document}
\maketitle

\vfill

\newpage

\setlength{\parindent}{0pt}
\section{Introduction}

The framework of any holographic construction elementally relies on the matching of symmetries on both sides of the duality. In the case of the AdS/CFT correspondence, the bulk isometries act as the group of conformal isometries on the boundary of the spacetime \cite{tHooft:1993dmi,Susskind:1994vu,Maldacena:1997re,Witten:1998qj,Gubser:1998bc}. Following this matching, there is an extrapolate dictionary that allows to define the operators of a lower-dimensional CFT which lives on the boundary of AdS and captures the dynamics of the bulk theory. Although the boundary of Anti de-Sitter is timelike, the appearance of null hypersurfaces as the spacetime boundary is also quite frequent, the first $-$ and arguably, the most important $-$ example being flat spacetime. In the case of null boundaries, the intrinsic metric degenerates and the Lorentzian structure is lost. One can however define a structure equipped with a degenerate metric and its kernel vector field, which goes by the name of \textit{Carroll manifold}. On such a manifold, the symmetry algebra is modified to the so-called \textit{Carroll algebra}, which can alternatively be obtained as an İnönü-Wigner contraction of the Poincaré algebra \cite{LevyLeblond,SenGupta:1966qer}.\\

From the point of view of four-dimensional asymptotically-flat spacetimes, the conformal Carroll symmetries induced on such a codimension-one null hypersurface are naturally isomorphic to the infinite-dimensional $\mathfrak{bms}_{4}$ algebra\footnote{In the nomenclature used in the literature, we mean here the extended $\mathfrak{bms}_{4}$ algebra including supertranslations and superrotations \cite{Barnich:2009se, Barnich:2010eb}.}\cite{Duval_2014,Bagchi:2010zz}, with their global subalgebra corresponding to the İnönü-Wigner contraction of the relativistic Conformal algebra. This isomorphism led to the proposition of codimension-one Carrollian CFTs as duals to quantum gravity in flat spacetime. Alternatively, one can motivate this correspondence by limiting arguments. The Minkowski metric can be obtained by taking the large radius limit of AdS. This large radius ($\ell \to \infty$) contraction of AdS isometries amounts to the $c\to 0$ contraction of the conformal group acting on the boundary. Hence, taking the $\Lambda \to 0$ limit in the bulk dials the speed of light to zero on the boundary. From this line of arguments, a significant amount of literature aimed towards flat space holography emerged in the last decade, initially for three dimensions 
\cite{Bagchi:2010zz,Bagchi:2012cy,Barnich:2012xq,Bagchi:2012xr,Barnich:2012rz,Barnich:2013yka,Bagchi:2014iea,Hartong:2015usd,Jiang:2017ecm}, and more recently for both three and four dimensions in \cite{Bagchi:2016bcd,Ciambelli:2018wre,Donnay:2022aba,Bagchi:2022emh,Donnay:2022wvx,Bagchi:2023fbj,Nguyen:2023vfz,Mason:2023mti,Alday:2024yyj,Kraus:2024gso,Hao:2025btl}. The Carrollian ($c \to 0$) contraction of Lorentzian symmetries was actually first observed in the sixties. Although it was initially studied as a curious limit, the growing amount of literature confirms the importance of this symmetry group.
 The reason why these Carrollian symmetries appear on so many occasions is its inherent relation with null hypersurfaces. As the Carrollian structures arise on the intrinsic geometries of null hypersurfaces embedded in higher-dimensional Lorentzian spacetimes, the degrees of freedom that live on these null hypersurfaces are naturally Carrollian. Apart from the context of flat space holography, Carrollian field theories have found applications in many other context, ranging from condensed matter \cite{Bagchi:2022eui,Bidussi:2021nmp,Kasikci:2023zdn,Figueroa-OFarrill:2023vbj,Figueroa-OFarrill:2023qty} to cosmology \cite{deBoer:2021jej,deBoer:2023fnj} and string theory \cite{Bagchi:2013bga,Bagchi:2015nca,Bagchi:2021ban}.\\ 
 
 The {\itshape isotropic} conformal extension of Carrollian symmetries, obtained by solving the conformal Carroll Killing equations differs greatly from the relativistic conformal group, as the conformal Carroll group is infinite-dimensional, even in $d>2$. Particularly, in three dimensions, this infinite-dimensional group consists of the local conformal transformations of the two-dimensional spatial slice along with arbitrary space-dependent time translations, \textit{i.e.} the four-dimensional BMS group. In the rest of this paper, we will refer to this group as the $z=1$ \textit{conformal Carroll group}, whose ten-dimensional global finite subgroup is the İnönü-Wigner contraction of the three-dimensional conformal group $SO(2,3)$. 
 As will be discussed in details below, the group contains a dilation generator $D = t \partial_t + x^i \partial_i$ scaling space and time {\itshape isotropically}. This can be generalised to
 {\itshape anisotropic} conformal Carroll symmetries 
 and shown to include in particular an anisotropic scaling generated by $D^{(z)} = z t \partial_t + x^i \partial_i$, similarly to Lifshitz-type theories \cite{Kachru:2008yh}. While the latter theories do not include Lorentzian boosts, theories with those symmetries include Carrollian boosts instead, leading to so-called \textit{anisotropic Conformal Carroll} theories with symmetry algebra denoted
$\mathfrak{ccar}_3^{(z)}$ (including in particular $t \to \lambda^z t, \, \,x^i \to \lambda x^i$). Interestingly, all of these conformal versions, labelled by a real number $z$, are infinite-dimensional  \cite{Bekaert:2024itn}.\\
 
 In this paper, we will focus on anisotropic Conformal Carroll theories in two and three dimensions. Their symmetry algebras for any $z$ consist of supertranslations and superrotations, but the Virasoro weights of the supertranslation generators change depending on $z$. As mentioned, the conformal Carroll symmetries acting on the boundary of Minkowski space correspond $z=1$ (and hence contain isotropic scalings consistent with Lorentz symmetry). Given the fact that $z=1$ Carrollian CFTs are putative duals of gravity in asymptotically-flat spacetimes, one may be interested in knowing if an analogue correspondence exists for generic $z$ Carroll CFTs. In particular, to establish the zeroth level of such a correspondence, one should answer the following question: is there any bulk metric in one higher dimension that realises arbitrary-$z$ global conformal Carroll symmetries as background isometries, and more preferably, is there a phase space realising the whole set of infinite-dimensional symmetries? Holography for non-relativistic and Lifshitz field theories are not a new subject by any means. They originated from the classic works \cite{Son:2008ye,BalaMcGreevy,Kachru:2008yh,Taylor:2008tg,Dong:2012se}, which identify bulk spacetimes with the necessary symmetries. A vast amount of literature subsequently followed towards establishing a holographic dictionary. In this article, we take similar steps,
 focusing on anisotropic Carroll symmetries. \\

We will see that the relevant family of metrics falls into the class of plane wave spacetimes $-$ or more generally \textit{plane-fronted waves with parallel rays} (pp-waves). Such spacetimes are defined as metrics that admit a covariantly-constant null Killing vector. The most general pp-wave metric can be expressed in Brinkmann coordinates as
\begin{align}
	ds^2 = 2dx^+dx^-+K(x^+,x^a)du^2+2A_a(x^+,x^a)dx^+dx^a+d\vec{x}^{\hspace{1pt}2}.
\end{align}
Plane waves are a special class of pp-wave metrics having $A_a(x^+,x^a)=0$ and \linebreak ${K(x^+,x^a)=A_{ab}(x^+)x^ax^b}$. 
A connection between plane waves and Carroll symmetries has been observed in \cite{Duval:2017els}. In particular, the five isometries of four-dimensional generic plane waves were shown to be identified with three-dimensional Carrollian symmetries without rotation. \\

A preliminary step for a holographic description relating these metrics and anisotropic conformal Carroll field theories is to identify a vacuum metric realising the global \textit{conformal} symmetries at the boundary. In the same vein as the Minkowski metric realises Poincaré ($z=1$ conformal Carroll) symmetries, we will highlight a four-dimensional plane wave metric realizing the global symmetries of $\mathfrak{ccar}_3^{(0)}$, given by
\begin{align}
\label{NWlike}
	ds^2=2dx^+dx^-+\frac{1}{4}\vec{x}{\,^2}(dx^+)^2+d\vec{x}{\,^2}.
\end{align}
This metric can be recognised as a double analytic continuation of the Nappi-Witten background \cite{Nappi:1993ie}, and we will present a generalisation for any $z$ later on. A feature of all plane wave metrics present in this work is that their null boundary is of co-dimension one. This is different from the typical plane wave metrics obtained in the BMN limit of AdS/CFT \cite{Berenstein:2002jq} as Penrose limits of AdS$_p \times S^q$ spaces \cite{Marolf_2002, Hubeny_2002}.
Similarly to flat space, the isometries, when projected on this null boundary, induce an anisotropic conformal Carroll structure on its intrinsic Carroll geometry. \\

\textbf{Outline of the paper}

The paper is organised as follows. Section \ref{sec2} provides a review of Carrollian geometry and the various conformal extensions of Carroll symmetries, with a particular focus on the algebraic structure for arbitrary scaling exponent $z$. In Section \ref{sec3}, we study the anisotropic conformal Carroll field theory, by turning to the construction of field theories invariant under these symmetries, concentrating on the case $z=0$ in both two and three dimensions. These theories share the same infinite-dimensional symmetry algebra as Warped CFTs, though we emphasize the differences stemming from their distinct global subgroups. We examine the Carrollian stress tensor, derive its transformation rules under symmetry generators, and compute correlation functions, including subtleties related to vacuum structure and normal ordering. In Section \ref{bulk}, we investigate the bulk realisation of these symmetries, identifying a specific plane wave spacetimes presenting global $z=0$ conformal Carroll symmetries. After gauge fixing, we propose a phase space whose residual diffeomorphisms reproduce the $z=0$ conformal Carroll algebra. Then, we extend our results to arbitrary $z$ and discuss the conformal boundary of the plane wave spacetimes we consider.

\newpage

\bigskip
\setlength{\parindent}{0pt}
\section{Conformal Carroll symmetries} \label{sec2}

 The Carroll group refers to the ultra-local limit of the Poincaré group, and can be implemented by taking the speed of light to zero. This amounts to scaling the time direction with a small parameter $\varepsilon$ and keeping spatial directions untouched: $t \to \varepsilon t$ and $x^i \to x^i$. At the level of the symmetry generators, this results in a so-called \textit{İnönü-Wigner contraction}, which implies an appropriate rescaling of the generators. By implementing this procedure explicitly on the Lorentz generators, one ends up with the following Carroll generators: 
\begin{align}
	H=\p_t, \quad P_i=\p_i, \quad J_{ij}=x_i\p_j-x^j\p_i, \quad C_i=x_i\p_t,
\end{align}
which span the Carroll algebra
\begin{equation}
\begin{aligned}
	&[J_{ij},P_k]=\delta_{ik}P_j-\delta_{jk}P_i, \qquad 
	[J_{ij},C_k]=\delta_{ik}C_j-\delta_{jk}C_i, \\ 
	&[J_{ij},J_{kl}]=SO(d), \qquad \qquad \; \; \;
	[P_i,C_j]=\delta_{ij}H.
\end{aligned}
\end{equation}
In $(d+1)$-dimensional spacetime, $i=1,\dots,d$ and $J_{ij},C_k,P_i$ and $H$ respectively denote spatial rotations, Carroll boosts, spatial momenta and Hamiltonian. Notice that the boosts now commute, which emphasises the non-relativistic nature of the Carroll symmetry, and the Hamiltonian $H$ enters the algebra as a central element only.\\

  One can define Carroll symmetry from a non-limiting, geometrical perspective. To do so, one introduces \textit{Carrollian manifolds} \cite{Henneaux:1979vn}, which are smooth manifolds endowed with a degenerate two-tensor field $h_{\mu \nu}$ and a nowhere-vanishing kernel vector field $\tau^\mu$. These objects $\tau^\mu$ and $h_{\mu\nu}$ are invariant under the group of local symmetries, namely the \textit{Carroll group}, thus providing the notion of Carrollian metric. For flat Carrollian structures, one has
\begin{align}
	\tau^\mu=(1,0^i), \quad h_{\mu\nu}=\mathrm{diag} (0,\delta_{ij}).
\end{align}
In this case, the Carroll curvature\footnote{We direct the interested reader to \cite{Hartong:2015xda} for gauged Carroll algebras and Carroll curvature.} can be shown to vanish and the isometries coincide with the Carroll group. As addressed previously, the degenerate structure on Carroll manifolds allows to scale the space and time directions differently, resulting in various conformal extensions of Carrollian isometries. This is reflected in the conformal isometry equations \cite{Duval:2014uva}
\begin{align} 
\label{Confcarr}
	\mathcal{L}_{\xi} \tau^\mu=-\frac{\lambda (x)}{k}\tau^\mu, \quad \mathcal{L}_{\xi} h_{\mu \nu}= \lambda (x) h_{\mu \nu}.
\end{align}
In the above equations, the parameter $k$, called the \textit{level}, captures the relative scaling between space and time components. It is useful to introduce the scaling exponent $z\equiv 2/k$. The specific value $k=2$, correspondingly $z=1$, represents a scenario where space and time are homogeneously scaled (notice that $h_{\mu \nu}$ is quadratic in the vielbein). For flat Carrollian manifolds and $k=2$, Eq. \eqref{Confcarr} results in an infinite-dimensional symmetry group that is isomorphic to the BMS group in one higher dimension. We will primarily focus on the $d=2$ and $d=3$ cases. In two dimensions, the $z$-conformal Carroll equations are solved by
\begin{align}
L_n=x^{n+1}\p_x+z(n+1)x^nt\p_t,\quad M_n=x^{n}\p_t,
\end{align}
which satisfy the following algebra \cite{Farahmand_Parsa_2019}: 
\begin{align} 
\label{z-bms3}
    [L_n,L_m]=(n-m)L_{m+n}, \quad [L_n,M_m]=((n+1)z-m)M_{m+n}, \quad [M_n,M_m]=0.
\end{align} 
It is also important to point out that, owing to the absence of spatial rotation in two dimensions, the two-dimensional Carroll and Galilei groups are isomorphic. Thus, the above algebra is isomorphic to the two-dimensional anisotropic Galilean conformal algebra \cite{Chen:2019hbj}. 
Notice that for $z=1$, this is nothing other than the well-known BMS$_3$ algebra which is recognized after shifting $m \rightarrow m+1$ and renaming $M_m \rightarrow M_{m-1}$\cite{Afshar:2024llh}.\\

  Our primary interest is the $z=0$, that is $k\to \infty$, conformal Carroll algebra. This symmetry algebra is a special example, as it can be realised in three-dimensional topologically massive gravity with certain boundary conditions, namely in asymptotically Warped AdS$_3$ (WAdS$_3$) \cite{
Compere:2009zj, Compere:2007in, Compere:2008cv, Blagojevic:2009ek, Henneaux:2011hv, Anninos:2010pm} or asymptotically Warped flat \cite{Detournay:2019xgl} spacetimes. The spacelike WAdS$_3$ metric is a deformation of global AdS$_3$ that breaks the $SL(2,\mathbb{C})$ symmetry to $SL(2,\mathbb{R})\times U(1)$ \cite{Anninos:2008fx, Israel04, Rooman_1998}. This metric appears universally in the near-horizon geometry of four-dimensional extremal Kerr black holes at a fixed polar angle. Asymptotically-WAdS$_3$ spaces enjoy the richness of three-dimensional gravity in AdS$_3$ and the phase space also contains interesting black hole solutions, called \textit{WAdS$_3$ black holes} \cite{Bouchareb_2007,YNutku_1993,Gurses:1994bjn, Anninos:2008fx}. These various considerations have led to a holographic description of Warped AdS spacetimes in terms of dual two-dimensional Warped CFTs \cite{Hofman:2011zj, Detournay:2012pc}, respecting the same infinite-dimensional $z=0$ algebra given in Eq. \eqref{z-bms3}. However, there are subtle differences between Warped CFTs and $z=0$ Carrollian CFTs, although they share the same symmetries, mainly stemming from the fact that the global $SL(2,\mathbb{R})\times U(1)$ does not contain the two-dimensional Carroll group. Since a Carrollian field theory is defined to have a  Carroll-invariant vacuum state, the Warped CFT vacuum is not Carrollian. Instead, the algebra (\ref{z-bms3}) admits 
  another four-parameter global subgroup spanned by the generators $\{L_0,L_{-1},M_{0},M_{1}\}$\footnote{Notice that these are precisely the global symmetries of the Warped flat spacetime \cite{Anninos:2008fx, Detournay:2019xgl}}. These generators represent the Carroll generators, along with the spatial dilation $z=0$. More explicitly, one can identify $L_0$ with the $z=0$ dilation, $L_{-1}\equiv P_1$ the spatial momentum, $M_0\equiv H$ the Hamiltonian and $M_1\equiv C_1$ the Carroll boost. In a field theory, these two different symmetry groups imply two different normal orderings. Hence, they give rise to two different spectra around these vacua, which distinguish Warped CFTs from Carrollian ones. In the next section, we will return to this discussion as we will compute their associated correlation functions.\\

  In three dimensions, for generic $k$, the above conformal isometry equation is solved by
\begin{align} \label{confkill}
	\xi=\left(\alpha(z,\z)+\frac{2}{k}((\p_z f^z(z)+\p_{\z}f^{\z}(\z))t \right) \p_t+f^z(z)\p_z+f^{\z}(\z)\p_{\z},
\end{align}
where we have used complex coordinates $z,\bar{z}$ for spatial directions. As advertised in the introduction, these conformal Killing vectors are parametrised by three functions $\alpha(z,\bar{z})$, $f^z(z)$ and $f^{\bar{z}}(\bar{z})$, which respectively generate space-dependent time translations and conformal transformations on two-dimensional spatial slices\footnote{The Carrollian structures defined by  $\tau^\mu$ and $h_{\mu\nu}$ are preserved not only by the finite-dimensional Carroll group, but also by an infinite set of supertranslations. In $(2+1)$-dimensions, corresponding to a two-dimensional spatial slice, these symmetries can be further extended by superrotations, which act as conformal isometries. For $d \geq 4$, the conformal isometries of a Carrollian manifold reduce to supertranslations combined with the finite-dimensional conformal symmetries of a $(d-1)$-dimensional space \cite{Duval:2017els}.
}. When mode-expanded, this conformal Killing vector splits into the following generators:
\begin{align} \label{z-killing}
	\hspace{-5 pt}L_n=-z^{n+1}\partial_z+\frac{2}{k}(n+1)z^n t\p_t, \quad \bar{L}_n=-\bar{z}^{n+1}\partial_{\bar{z}}+\frac{2}{k}(n+1)\z^nt\p_t, \quad M_{r,s}=z^r\bar{z}^s\partial_t.
\end{align}
They satisfy the following algebra
\begin{equation} \label{ccarr_z}
\begin{aligned}
	[L_n,L_m]&=(n-m)L_{n+m}, \hspace{1.5cm}\quad [\bar{L}_n,\bar{L}_m]=(n-m)\bar{L}_{n+m} \\ 
	[L_n,M_{r,s}]&=\left(z\frac{n+1}{2}-r\right)M_{r+n,s}, \quad [\bar{L}_n,M_{r,s}]=\left(z\frac{n+1}{2}-s\right)M_{r,s+n} \\ 
	[M_{r,s},M_{p,q}]&=0,
\end{aligned}
\end{equation}
to which we refer as the $d=3$, $z=2/k$ conformal Carroll algebra.
\noindent For $z=1$, it is indeed the extended BMS$_4$ algebra, with  
$\{L_{0},L_{\pm1},\bar{L}_0, \bar{L}_{\pm1}, M_{0,0},M_{0,1},M_{1,0},M_{1,1}\}$  
representing the generators of the ten-dimensional Poincaré subalgebra.\\

  As we have already mentioned, we will mostly focus on the $z=0$ conformal Carroll algebra. This algebra has previously appeared in the context of near-horizon symmetries in \cite{Donnay:2015abr,Grumiller:2019fmp,Afshar:2016wfy}. There are infinitely-many finite-dimensional subalgebras of the $z=0$ algebra, see for example the discussion in Appendix \ref{appendix}. There are three seven-dimensional subalgebras of particular interest:
\begin{itemize}
	\item $SL(2, \mathbb{R}) \times SL(2, \mathbb{R}) \times U(1)$, spanned by $\{L_{0},L_{\pm1},\bar{L}_{0},\bar{L}_{\pm1},M_{0,0}\}$ 
    \begin{equation}
	\begin{aligned}
		[L_0,L_1]=-L_1, \quad [L_0,L_{-1}]=L_{-1} \\ 
		[\bar{L}_0,\bar{L}_1]=-\bar{L}_1, \quad [\bar{L}_0,\bar{L}_{-1}]=\bar{L}_{-1}  
	\end{aligned}
    \end{equation}
	\item The type-D, $z=0$ conformal Carroll algebra: $\{L_0,\bar{L}_0,L_{-1},\bar{L}_{-1},M_{0,0},M_{1,0},M_{0,1}\}$ algebra \cite{hofman2015warped, Afshar:2024llh}, spanned by
    \begin{equation}
	\begin{aligned}
		[L_0,L_{-1}]&=L_{-1},  \; \quad [L_{-1},M_{1,0}]=-M_{0,0}, \\ 
		[\bar{L}_0,\bar{L}_{-1}]&=\bar{L}_{-1}, \quad \; [\bar{L}_{-1},M_{0,1}]=-M_{0,0} \\ 
		[L_{0},M_{1,0}]&=-M_{1,0}, \quad \! [\bar{L}_0,M_{0,1}]=-M_{0,1} \label{typeD3d}
	\end{aligned}
    \end{equation}
   \item The type-K, $z=0$ conformal Carroll algebra, spanned by $\{J\equiv L_0-\bar{L}_0,L_{-1},\bar{L}_{-1},M_{0,0},$
   $M_{1,0},M_{0,1}, M_{1,1}\}$ algebra \cite{Afshar:2024llh}:
   \begin{equation}
   \begin{aligned}
		[J,L_{-1}]&=L_{-1}, \quad \; \; \; [L_{-1},M_{1,0}]=-M_{0,0},\qquad [L_{-1}, M_{1,1}]=-M_{0,1}, \\ 
		[J,\bar{L}_{-1}]&=-\bar{L}_{-1}, \quad [\bar{L}_{-1},M_{0,1}]=-M_{0,0}, \qquad [\bar{L}_{-1}, M_{1,1}]=-M_{1,0}, \\ 
		[J,M_{1,0}]&=-M_{1,0}, \qquad \hspace{-0.3 pt}[J,M_{0,1}]=M_{0,1}.
	\end{aligned}
    \end{equation}
	
    \end{itemize}
The first case is perhaps an obvious one, as it contains the global conformal symmetries of the two-dimensional spatial slice and time translations, but it does not contain the Carroll boost generators. More interestingly, the $z=0$ type-D and type-K conformal Carroll algebras contain all the Carroll symmetries along with the $z=0$, \textit{i.e} spatial, dilation or the $M_{11}$ special conformal transformation, respectively. One can easily be convinced that these are the only two seven-dimensional algebras that contain the full Carroll algebra. The reason for our discussion of seven-dimensional subalgebras will be made explicit in Section \ref{bulk}, but let us already comment on it: in the spirit of holography, we will be looking at spacetimes that realise conformal extensions of the Carroll group as their isometry group. It has been shown in \cite{Duval:2017els} that generic four-dimensional plane waves naturally include the three-dimensional Carroll group without rotations as part of their isometry group. For conformally-flat plane waves, the isometry group contains the full three-dimensional Carroll group. In Section \ref{bulk}, we will be looking to realise $d=3$ conformal Carroll symmetries in the bulk, by identifying vacuum metrics exhibiting a global subgroup of the conformal Carroll symmetries as isometries, as well as a family of metrics whose asymptotic symmetries coincide with conformal Carroll. We will also briefly discuss the lower-dimensional case. As will be explained later, plane waves in four dimensions have at most (except for the trivial case of flat space where it jumps to ten with no possible in-between) seven isometries and it will thus not be possible to realise larger finite-dimensional subalgebras, such as the D-K or D-K-K$_i$ algebras of \cite{Afshar:2024llh} as exact isometries of such spacetimes (but will be realized as asymptotic symmetries). For four-dimensional spacetimes, nine-dimensional \textit{global} isometry algebras are not permitted \cite{hall2004symmetries}. Eight-dimensional global isometry algebras can occur, in which case the manifold is of constant curvature and admits, locally, a ten-dimensional algebra. This \textit{a priori} does not forbid a four-dimensional spacetime to admit an eight- or nine-dimensional local isometry algebra, 
but we have not been able to find such spacetimes (which would also have to contain Carroll symmetries).

\section{Aspects of Carrollian CFTs} \label{sec3}

In this section, we turn our attention to the field theories that preserve infinite-dimensional conformal Carroll symmetries, mainly focusing on $z=0$. As addressed in the introduction, the infinite-dimensional extension of conformal symmetries is natural in the Carrollian regime, even in spacetimes of dimensions larger than three. This infinite enhancement of symmetries makes the investigation of these field theories interesting in its own right. A considerable amount of literature exists in two dimensions regarding Warped CFTs (WCFTs) \cite{Hofman:2011zj,Detournay:2012pc} sharing the symmetry algebra of $\mathfrak{ccar}_2^{(0)}$, \textit{i.e.} that of $z=0$ two-dimensional Carrollian CFTs (but differing in their global symmetry group, the vacuum of WCFTs not being invariant under the Carroll boost). We now investigate some properties of these field theories, in particular their stress tensor and correlation functions, in two and three dimensions.

\subsection{Carrollian stress tensor and transformation laws}

\subsubsection{In $d=3$}

The currents associated with $z=0$ conformal Carroll symmetries can be constructed by the standard method of contracting the associated Killing vectors fields \eqref{confkill} with the stress tensors
\begin{align} \label{current}
J^{\alpha}=T^{\alpha}_{\beta}\xi^{\beta}.
\end{align}
Stress tensors on Carrollian manifolds can be defined by the variations of an action with respect to Carrollian geometric elements \cite{Baiguera:2022lsw}. On Carrollian manifolds, the tangent space symmetries include Carroll boosts and rotations. These local Carroll symmetries dictate the symmetry properties of Carrollian stress tensors. These are given by
\begin{align}
    T^u_i=0, \quad T^i_j=T^j_i.
\end{align}
They are analogous to the symmetric stress tensor condition of relativistic theories, which follows from the local Lorentz invariance. The skew-symmetric first condition is a consequence of the off-diagonal nature of Carroll boosts and implies the vanishing of the energy flux. Furthermore, $z=0$ Weyl symmetry makes the spatial components traceless\footnote{It is straightforward to verify that the analogous condition of the $z$-Weyl invariant theory would be $zT^u_u+T^i_i=0$.}
\begin{align}
    T^i_i=0.
\end{align}
Using these conditions, as well as the expression of the Killing vector $\xi^\mu$, one can verify that the current \eqref{current} is conserved. Subsequently, the charges can be expressed as integral of $J^u$ over the spatial slice as
\begin{align} \label{charges}
	Q_{\xi}&=\int d^2z\, J^u (z,\z) \\ \nonumber
    &=\int d^2z  \,
\big(T^u_u(z,\z)\alpha(z,\z)+T^u_z(z,\z)f^z(z)+T^u_{\z}f^{\z}(\z)\big),
\end{align}
 where $\alpha(z,\z),f^z(z)$ and $f^{\z}(\z)$ are parametrising functions associated with supertranslations and superrotations, respectively. For the ease of notation, we shall denote these stress tensor components respectively by $T^u_u \equiv M(z,\z)$, $T^u_z \equiv T_z(z,\z)$ and $T^u_{\z} \equiv T_{\z}(z,\z)$. Upon expanding these parameters in modes
\begin{align}
\alpha(z,\z)=\sum_{r,s}\alpha_{rs}z^r\z^s, \quad f^z(z)=\sum_n a_nz^{n+1}, \quad f^{\z}(\z)=\sum_{n} \bar{a}_n \z^{n+1},
\end{align}
we can express the charges \eqref{charges} in the standard notation as
\begin{align}
Q_{\xi}=\sum_{r,s}\alpha_{rs}M_{r,s}+\sum_n a_nL_n+\sum_{n} \bar{a}_n \bar{L}_n,
\end{align}
where
\begin{align}
	L_n=\int d^2z \, T_z(z,\z)z^{n+1}, \quad \bar{L}_{n}=\int d^2z  \, T_\z(z,\z)\z^{n+1}, \quad M_{r,s}=\int d^2z \, M(z,\z)z^r\z^s.
\end{align}
 The algebra of the charges dictates how these operators transform under these transformations. Subsequently, we shall assume that the charges satisfy the same algebra as the Killing vectors and the transformation properties of  $M(z,\z)$, $T_z(z,\z)$ and $T_{\z}(z,\z)$. A similar analysis for BMS$_4$ invariant field theories was carried out in \cite{Dutta:2022vkg}. Here we repeat the procedure generalizing for arbitrary $z$. \\
 
Let us first put our attention on the Virasoro subalgebra. In a two-dimensional CFT, the Virasoro algebra dictates the transformation of holomorphic and antiholomorphic stress tensor components. They transform as primaries with conformal weights $(2, 0)$ and $(0, 2)$. In a similar fashion, the underlying infinite-dimensional algebra of conformal Carrollian symmetries also confirms that $T_z(z,\z)$,  $T_{\z}(z,\z)$ and $M(z,\z)$  transform as primaries. Their conformal weights are identified as follows: let us first consider $T_z(z,\z)$, \textit{i.e.}, the generator of $L_n$. The left-moving Virasoro algebra fixes the holomorphic weight $h=2$, just as in a two-dimensional CFT. However, in contrast to two-dimensional CFTs, $[\bar{L}_n,L_m]=0$ implies
\begin{equation}
\begin{aligned}
	&\left[\bar{L}_n,\int d^2z \, T_z(z,\z)z^{m+1}\right]=0 \\ \nonumber
	\Rightarrow \quad	&[\bar{L}_n, T_z(z,\z)]=z^{n+1}\p_zT_z(z,\z)+(n+1)T_z(z,\z)z^n.
\end{aligned}
\end{equation}
 Using the same set of commutators, it is straightforward to show that $T_\z(z,\z)$ also transforms as a Virasoro primary with conformal weights (1,2). The same goes for $M(z,\z)$, where the commutation relations imply weights $(1,1)$. To summarise, it is possible to define local operators $T_z(z,\z)$, $T_{\z}(z,\z)$ and $M(z,\z)$ using the stress tensors of a three-dimensional Carroll CFT. They transform as conformal primaries with the Virasoro weights listed in Table \ref{t1}.

\begin{table}[h!]
	\centering
	\begin{tabular}{||c | c c c||} 
		\hline
		Weights &  $M$ & $T_z$ & $T_{\bar{z}}$ \\ [0.75ex] 
		\hline\hline
		$h$ &  1 & 2 & 1 \\ 
		\hline
		$\bar{h}$ & 1 & 1 & 2 \\
		\hline
	\end{tabular}
	\caption{Conformal weights of the local operators in $d=3,\, z=0$ case.}
	\label{t1}
\end{table}
\medskip
One can repeat these arguments for a generic $z$-Carroll CFT. In that case, one obtains the weights of Table \ref{t2}.
\begin{table}[h!]
	\centering
	\begin{tabular}{||c | c c c||} 
		\hline
		Weights &  $M$ & $T_z$ & $T_{\bar{z}}$ \\ [0.75ex] 
		\hline\hline
		$h$ & $ 1+z/2$ & 2 & 1 \\ 
		\hline
		$\bar{h}$ & $1+z/2$ & 1 & 2 \\
		\hline
	\end{tabular}
	\caption{Conformal weights of the local operators in $d=3$, generic $z$ case.}
	\label{t2}
\end{table}

\paragraph{Transformation with respect to supertranslations} 
\smallskip

The transformation of $M(z,\z)$, $T_z(z,\z)$ and $T_{\z}(z,\z)$ can, again, be figured out using the commutators between $L_n$ and $M_{r,s}$. For instance,
\begin{align}
	&[M_{r,s},L_n]=rM_{n+r,s}  \\ \nonumber
	& \hspace{-4 pt}\left[M_{r,s},\int d^2z \, T_z(z,\z)z^{n+1}\right]=r\int d^2z \, M(z,\z)z^{n+r}\z^s  \\ \nonumber
	\Rightarrow \quad \, &[M_{r,s},T_z(z,\z)]=rM(z,\z)z^r\z^s.
\end{align}
Generically, for any arbitrary function $\alpha(z,\z)$, we have 
\begin{equation}
	\delta_{\alpha(z,\z)}T_z(z,\z)=\partial_z\big(\alpha(z,\z)M(z,\z)\big), \quad \delta_{\alpha(z,\z)}T_\z(z,\z)=\partial_\z\big(\alpha(z,\z)M(z,\z)\big).
\end{equation}
As supertranslations form an abelian subalgebra, one has
\begin{equation}	\delta_{\alpha(z,\z)}M(z,\z)=0.
\end{equation}
The invariance of $M(z,\z)$ under supertranslations is irrespective of the scaling exponent $z$. However, $T_z(z,\z)$ and $T_\z(z,\z)$ transform as 
\begin{align}
 	&\delta_{\alpha(z,\z)}T_z(z,\z)=\partial_z\alpha(z,\z)M(z,\z)+\left(\frac{z}{2}+1\right)\p_zM(z,\z)\alpha(z,\z), \\ \nonumber
    &\delta_{\alpha(z,\z)}T_\z(z,\z)=\partial_\z\alpha(z,\z)M(z,\z)+\left(\frac{z}{2}+1\right)\p_{\z}M(z,\z)\alpha(z,\z), \\ \nonumber
    &\delta_{\alpha(z,\z)}M(z,\z)=0.    
\end{align}

\subsubsection{In $d=2$}

In two dimensions, the analogous symmetry algebra is known as the Warped Conformal algebra and is given by the following commutation relations \cite{Detournay:2012pc} 
\begin{align}
    [L_m,L_n] &= (m-n)L_{m+n} + \frac{c}{12} m(m^2-1)\delta_{m+n,0},\\
    [L_m,M_n] &= -nM_{n+m}-ik'(m^2-m)\delta_{m+n,0},\\
    [M_m,M_n] &= \frac k2 m \delta_{m+n,0}.
\end{align}
In the literature, field theories invariant under this symmetry group, \textit{i.e.} the Warped CFTs, are usually viewed as a chiral CFT with an additional U(1) Kac-Moody current. However, at classical level, these field theories can alternatively be viewed as $z=0$ Carrollian CFTs sharing the same symmetry algebra, but a different global subgroup as WCFTs, now $\{L_0, M_0, L_{-1},M_{1}\}$. From this point of view, the generators of the symmetry algebra could also be obtained from a two-dimensional Carroll stress tensor. This analysis is almost identical to that of previous section. Here we briefly summarise the main result for the sake of completeness. We demand the action is invariant under the Local Carroll boost and the $z=0$ Weyl symmetry. These conditions imply
\begin{align}
    T^{x}_u=0, \quad T^{x}_{x}=0
\end{align}
Then, the generators $L_n$ and $M_n$ could be expressed as modes of the two other components,
\begin{align}
    L_n = -\frac{i}{2\pi} \int dx \, x^{n+1}T^u_x(x), \quad M_n = -\frac{1}{2\pi}\int dx \,x^nT^u_u(x).
\end{align}
 The infinitesimal transformations of the stress tensor components are given by
\begin{align}
    \delta_\varepsilon T(z) &= -\varepsilon(z)\partial_z T(z) - 2 \partial_z \varepsilon(z)T(z)-\frac{c}{12}\partial_z^3\varepsilon(z),\\
    \delta_\gamma T(z) &= - \partial_z \gamma(z)M(z)+k'\partial_z^2\gamma(z),\\
    \delta_\varepsilon M(z) &= - \varepsilon(z)\partial_z M(z) - \partial_z \varepsilon(z) M(z)-k'\partial_z^2\varepsilon(z),\\
    \delta_\gamma M(z)&= \frac k2 \partial_z \gamma(z),
\end{align}
where $T(z)$ is the right-moving energy momentum tensor, $M(z)$ a right-moving $\hat{\mathfrak{u}}(1)$ Kac-Moody current and we define
\begin{equation}   \delta_{\varepsilon+\gamma}=\delta_\varepsilon + \delta_\gamma = -i[T_\varepsilon, \cdot\,]-i[M_\gamma, \cdot \,],
\end{equation}
with
\begin{equation}
    T_\varepsilon = -\frac{1}{2\pi} \oint dz\, \varepsilon(z) T(z), \quad M_\gamma = -\frac{1}{2\pi} \oint dz\, \gamma(z) M(z).
\end{equation}

\subsection{Correlation functions}
\label{Sec:CorrFct}
The rest of this section is dedicated to fixing the correlation functions using symmetry arguments. We have already discussed the subtleties regarding vacuum states for these Carrollian CFTs. Here we shall solve the Ward identities associated with the two sets of generators and derive the two-point functions for both $d=2$ and $d=3$. Some results in this section were recently obtained in \cite{Afshar:2024llh}
with slight variations, which we highlight in the text.

\subsubsection{In $d=3$}

In three dimensions, the maximal set of commuting charges are given by the set $\{L_0,\bar{L}_0,M_{0,0}\}$. We define primary fields by labelling them by the eigenvalues of this set\footnote{Notice that in addition to $L_0$ and $\bar{L}_0$ we also label the operators with the eigenvalue of $M_{00}$. This is possible only for $z=0$ case, where all three of them commute. However if $z \neq 0$, it is only possible to label with two quantum numbers, as done in \cite{Afshar:2024llh}.} such that
\begin{align} \label{qnum}
	[L_0,\Phi(0)]=h\Phi(0), \quad [\bar{L}_0,\Phi(0)]=\h\Phi(0), \quad [M_{0,0},\Phi(0)]=Q\Phi(0),
\end{align}
with $Q$ the eigenvalue of the Hamiltonian, denoting the energy of the associated operator. Conformal dimensions $h$ and $\h$ can be traded for the scaling dimension $\Delta$ and spin $\sigma$ according to
\begin{align}
	\Delta=h+\h, \quad \sigma=h-\h.
\end{align}
The last equation in \eqref{qnum} essentially fixes the time evolution of the primary operators
\begin{align} \label{tevo}
[M_{0,0},\Phi(u,z,\z)]=i\p_u\Phi(u,z,\z)=Q\Phi(u,z,\z) \quad \Rightarrow \quad \, \Phi(u,z,\z)=e^{-iQu}\O(z,\z),
\end{align}
where we define the local operator $\O(z,\z)$ exclusively on the spatial slice at $u=0$. Subsequently, it can be shown that for higher supertranslations, we have
\begin{align}
	[M_{r,s},\Phi(u,z,\z)]=iz^r\z^s\p_u\Phi(u,z,\z)=Qz^r\z^s\O(u,z,\z).
\end{align}
The first two relations of \eqref{qnum}, supplemented by highest-weight conditions, lead to the following primary transformation rules exclusively for $\O(z,\z)$:
\begin{align}
	&	[L_n,\O(z,\z)]=z^{n+1}\p_z \O(z,\z)+(n+1)hz^n \O(z,\z),  \\ \nonumber
	&	[\bar{L}_n,\O(z,\z)]=\z^{n+1}\p_{\z} \O(z,\z)+(n+1)\h\z^n \O(z,\z).
\end{align}

Knowing these, we shall focus on their correlation functions. For the consistency of notation, we denote the two vacuum states\footnote{Type-K is not considered as it would yield the same result that will be obtained in Eq. \eqref{Eq:CorrFct}. The absence of dilation leaves the temporal dependency to be an arbitrary function of $u_1-u_2$, while the extra generator $M_{11}$ does not add any constraint.} as 
\begin{align} \label{VacWC}
	&\textbf{Vacuum W} & \{L_{0},L_{\pm1},\bar{L}_{0},\bar{L}_{\pm1},M_{0,0}\} |0 \rangle_I =0, \\ \nonumber
	&\textbf{Vacuum C} & \{L_{0},L_{-1},\bar{L}_{0},\bar{L}_{-1},M_{0,0},M_{0,1},M_{1,0}\} |0 \rangle_{II} =0,
\end{align}
where \textit{W} and \textit{C} refer to the $\SL \times \SL \times U(1)$ and Type-D Carroll invariant vacua, respectively. The two-point correlation functions of these primary operators are given by 
\begin{align}
	G^{(2)}(u_i,z_i,\z_i)=\langle \Phi_{\Delta_1,\sigma_1,Q_1}(u_1,z_1,\z_1)\Phi_{\Delta_2,\sigma_2,Q_2}(u_2,z_2,\z_2)\rangle.
\end{align}
Let us delve into each vacuum separately.\\

\paragraph{Vacuum W}
\smallskip
In this case, we need to solve the Ward identities associated to the symmetry generators $\{L_{0},L_{\pm1},\bar{L}_{0},\bar{L}_{\pm1},M_{0,0}\}$. The $M_{0,0}$ time translation Ward identity 
\begin{align}
	(\p_{u_1}+\p_{u_2}) G_W^{(2)}(u_i,z_i,\z_i)=0,
\end{align}
along with Eq. \eqref{tevo}, lead to conservation of energy
\begin{align}
	Q_1+Q_2=0.
\end{align}
This condition fixes the correlation function up to
\begin{align}\label{3dw}
	G_W^{(2)}(u_i,z_i,\z_i)=e^{-iQ_1(u_1-u_2)}\langle \O_{\Delta_1,\sigma_1}(z_1,\z_1)\O_{\Delta_2,\sigma_2}(z_2,\z_2)\rangle.
\end{align}
Furthermore, the other six generators $L_{0},L_{\pm1}$ and $\bar{L}_{0},\bar{L}_{\pm1}$ act exclusively on the spatial slice as global conformal transformations. The associated Ward identities determine the spatial part of the correlator, in the same way as in a two-dimensional CFT:
\begin{align}
	\langle \O_{\Delta_1,\sigma_1}(z_1,\z_1)\O_{\Delta_2,\sigma_2}(z_2,\z_2)\rangle= \frac{\delta_{\Delta_1,\Delta_2}\delta_{\sigma_1,\sigma_2}}{(z_1-z_2)^{2h}(\bar{z}_1-\bar{z}_2)^{2\bar{h}}}.
\end{align}
Finally, one obtains
\begin{align}
	G_W^{(2)}(u_i,z_i,\z_i)=\delta_{Q_1,-Q_2}\delta_{\Delta_1,\Delta_2}\delta_{\sigma_1,\sigma_2}\frac{e^{-iQ_1(u_1-u_2)}}{(z_1-z_2)^{2h}(\bar{z}_1-\bar{z}_2)^{2\bar{h}}}.
\end{align}
\medskip

\paragraph{Vacuum C} 
\smallskip
The Carroll boosts $\{M_{1,0},M_{0,1}\}$ replace the two special conformal transformations $\{L_1,\bar{L}_1\}$ and play a very special role. The spacetime translation Ward identity again fixes the correlation function up to
\begin{align}
	G_C^{(2)}(u_i,z_i,\z_i)=\delta_{Q_1,-Q_2}e^{-iQ_1(u_1-u_2)}f(z_{12},\z_{12}).
\end{align}
However, the boost Ward identities have non-trivial consequences. The Ward identity associated with $M_{1,0}$ is given by
\begin{align}
	&(z_1\p_{u_1}+z_2\p_{u_2})G_C^{(2)}(u_i,z_i,\z_i)=0 \\ \nonumber
	\Rightarrow \quad &iQ_1(z_1-z_2)f(z_1-z_2,\bar{z}_1-\bar{z}_2)=0.
\end{align}
The above equation has two solutions
\begin{equation}
\begin{cases}
    Q_1(z_1-z_2)=0, \\
    f(z_1-z_2,\bar{z}_1-\bar{z}_2)=\delta^{(2)}(z_1-z_2,\bar{z}_1-\bar{z}_2).
\end{cases}
\end{equation}
These solutions clearly classify two distinct scenarios, depending on the energy $Q_1$ of the associated operator. If $Q_1=0$, then the operators do not evolve in time and are thus functions of the spatial coordinates $z,\z$ only. These functions can further be constrained using the Ward identities associated with $L_0$ and $\bar{L}_0$. The solution is given by 
\begin{align}
	f(z_1-z_2,\bar{z}_1-\bar{z}_2)=\frac{1}{(z_1-z_2)^{h_1+h_2}(\bar{z}_1-\bar{z}_2)^{\bar{h}_1+\bar{h}_2}}.
\end{align}
 Thus, in this particular case, we have
\begin{align}
	G_C^{(2)}(u_i,z_i,\bar{z}_i)=\frac{1}{(z_1-z_2)^{h_1+h_2}(\bar{z}_1-\bar{z}_2)^{\bar{h}_1+\bar{h}_2}}.
\end{align}
Although the above expression shares some similarities with Eq. \eqref{3dw}, a crucial difference is that their correlator continues to be non-trivial for unequal values of scaling dimensions. \\

For $Q_1\neq0$, the operators evolve over time. Hence, the correlation functions also have a non-trivial time-dependence. However, the boost Ward identities allow only $\delta$-functions in the spatial sector:
\begin{align} \label{5.27}
	G_C^{(2)}(u_i,z_i,\bar{z}_i)=e^{-iQ_1(u_1-u_2)}\delta^{(2)}(z_1-z_2,\bar{z}_1-\bar{z}_2).
\end{align}
The presence of a $\delta$-function reflects the ultra-local nature of the correlator, which is typical in Carrollian theories \cite{Klauder:2000ud,SenGupta:1966qer,LevyLeblond,Henneaux:2021yzg,Banerjee:2020qjj}. It is not surprising that we encounter this branch in the correlation function for the Carrollian vacuum. As $\delta^{(2)}(z_1-z_2,\z_1-\z_2)$ scales inversely with the coordinates, this correlation is only non-trivial for operators with specific scaling dimension. This can be seen from $L_0$ and $\bar{L}_0$ Ward identities, given by
\begin{equation}
[(z_1\p_{z_1}+z_2\p_{z_2})+h_1+h_2]G^{2}(u_i,z_i,\z_i)=0.
\end{equation}
Plugging Eq. \eqref{5.27} in it, we end up with
\begin{equation}
	h_1+h_2=1.
\end{equation}
Similarly from the $\bar{L}_0$ Ward identity, we have 
\begin{equation}
	\h_1+\h_2=1.
\end{equation}
Finally, one obtains
\begin{equation}
	G_C^{(2)}(u_i,z_i,\bar{z}_i)=e^{-iQ_1(u_1-u_2)}\delta^{(2)}(z_1-z_2,\bar{z}_1-\bar{z}_2)\delta_{\Delta_1+\Delta_2,2}\delta_{\sigma_1,-\sigma_2} \delta_{Q_1,-Q_2}.
\end{equation}
Correlation functions for generic non-zero $z$ Carroll CFTs, recently derived in \cite{Afshar:2024llh}, are a generalisation of the $z=1$ case \cite{Bagchi:2022emh}. For completeness, we provide the expression
\begin{align} 
	G^{(2)}(u_i,z_i,\bar{z}_i)=(u_1-u_2)^{-(\Delta_1+\Delta_2-2)/z}\delta^{(2)}(z_1-z_2,\bar{z}_1-\bar{z}_2).
    \label{Eq:CorrFct}
\end{align}

\subsubsection{In $d=2$}
 As addressed previously, this infinite-dimensional algebra has two four-dimensional global subgroups. The previously well-studied subgroup  $SL(2,\mathbb{R}) \times U(1)$ is spanned by the generators $\{L_{0},L_{\pm 1},M_0 \}$. They generate two translations, a spatial dilation and a special conformal transformation. The other one is spanned by three Carroll generators along with a dilatation, \textit{i.e.} $\{L_{0},L_{-1},M_{0},M_1 \}$. Let us denote the associated vacuum states as 
\begin{align}
    &\textbf{Vacuum W} & \{ L_{0},L_{\pm1}, M_{0} \} |0 \rangle_W = 0, \\ 
    &\textbf{Vacuum C} & \{ L_{0}, L_{-1}, M_{0}, M_{1} \} |0 \rangle_C= 0. \label{VacuumC3D}
\end{align}
We shall again label the primary operators at origin using their eigenvalues. One has
\begin{align}
    [L_0,\Phi(0)]=\Delta \Phi(0), \quad [M_0,\Phi(0)]=Q\Phi(0).
\end{align}
Looking at these transformations at arbitrary points using the algebra yields
\begin{align}
&[L_n,\Phi(t,x)]=\Big(x^{n+1}\partial_x+(n+1)\Delta x^n \Big)\Phi(t,x),  \nonumber\\
&[M_n,\Phi(t,x)]=x^n\partial_t\Phi(t,x).
\end{align}
Notice that these fields can also labelled by the eigenvalue of the Hamiltonian $M_0$. This condition determines the time evolution of these fields as
\begin{equation}
    [M_0,\Phi(t,x)]=-i\partial_t \Phi(t,x)=Q \Phi(t,x) \quad
    \Rightarrow \quad \Phi(t,x)=e^{iQt}\O(x).
\end{equation}

\paragraph{Vacuum W}\smallskip
The correlation functions of $\O(x)$ entirely determines the correlator of $\Phi(t,x)$.  The above representation and correlation function in vacuum W has been studied previously in \cite{Song:2017czq} and is given by
\begin{align} \label{vacW}
    G_W^{(2)}(u,u',x,x')= \langle\Phi(u,x)\Phi(u',x')\rangle=e^{iQ_1(u_1-u_2)}\frac{1}{(x_1-x_2)^{2\Delta_1}}\delta_{Q_1+Q_2}\delta_{\Delta_1,\Delta_2}.
\end{align}

\paragraph{Vacuum C}\smallskip
The associated two-point function is determined by solving the Ward identities associated with $\{L_{0},L_{-1},M_{0,1} \}$. The time translation Ward identity reads 
\begin{align}
   &( \partial_{u_1}+ \partial_{u_2}) G_C^{(2)}(u_1,u_2,x_1,x_2)=0 \quad \Rightarrow \quad Q_1+Q_2=0,
\end{align}
thus giving
\begin{equation}
G^{(2)}_C(u_1,u_2,x_1,x_2)=e^{iQ_1(u_1-u_2)}\langle\O(x_1)\O(x_2)\rangle.
\end{equation}
Similarly the spatial translation invariance fixes the correlator to be
\begin{align}
(\partial_{x_1}+\partial_{x_2})G^{(2)}_C(u_1,u_2,x_1,x_2)=0 \quad \Rightarrow \quad \langle\O(x_1)\O(x_2)\rangle=f(x_1-x_2).
\end{align}
Furthermore, the Carroll boost Ward identity yields
\begin{align}
&(x_1\partial_{u_1}+x_2\partial_{u_2})G^{(2)}_C(u_1,u_2,x_1,x_2)=0,\nonumber\\ \nonumber
&Q_1(x_1-x_2)f(x_1-x_2)=0.
\end{align}
Typically, the above equation can be solved by either choosing $Q_1 = 0$ or setting \linebreak ${f(x_1 - x_2) = \delta(x_1 - x_2)}$. The requirement of dilation invariance further constrains these two branches differently. For  $Q_1 = 0$, the Ward identity is given by
\begin{align}
    \Big((x_1\partial_{x_1}+x_2\partial_{x_2})+(\Delta_1+\Delta_2)\Big)f(x_1-x_2)=0.
\end{align}
This equation is solved by
\begin{align}
    f(x_1-x_2)=\frac{1}{(x_1-x_2)^{\Delta_1+\Delta_2}}.
\end{align}
In this case, we have
\begin{align}
    G^{(2)}_C(u_1,u_2,x_1,x_2)=\frac{1}{(x_1-x_2)^{\Delta_1+\Delta_2}}.
\end{align}
On the other hand, if $f(x_1-x_2)=\delta(x_1-x_2)$, one obtains
\begin{align}
\Big((x_1\partial_{x_1}+x_2\partial_{x_2})+(\Delta_1+\Delta_2)\Big)\delta(x_1-x_2)=0\quad 
\Rightarrow \quad \Delta_1+\Delta_2=1
\end{align}
It follows that
\begin{equation}
  G^{(2)}_C(u_1,u_2,x_1,x_2)=e^{iQ_1(u_1-u_2)}\delta(x_1-x_2)\delta_{\Delta_1+\Delta_2,1}\delta
  _{Q_1+Q_2,0}.
\end{equation}
These two different branches of solutions clearly highlight different types of Carrollian transition amplitudes. A typical feature of a Carrollian theory is the vanishing of energy flux, indicating that particles with energy cannot travel in space. Spatial propagation is only possible if the particle carries no energy \cite{deBoer:2021jej}. The ultra-local and CFT-like branches reflect these two distinct scenarios in the Carrollian case. \\

It is possible to generalise this result for arbitrary value of the scaling exponent $z$. We recall that the algebra \eqref{z-bms3} is generated by the vector fields 
\begin{align}
L_n=x^{n+1}\p_x+z(n+1)x^nt\p_t, \quad M_n=x^{n}\p_t.
\end{align}
The main qualitative difference for $z\neq0$ is the fact the representations are not labelled by the eigenvalue of the Hamiltonian anymore. Hence, the time-dependent function does not assume the exponential form. This result has already been presented in \cite{Afshar:2024llh}: 
\begin{equation}
G^{(2)}(u_1,u_2,x_1,x_2)=(u_1-u_2)^{(\Delta_1+\Delta_2-1)/z}\delta(x_1-x_2)\delta
  _{Q_1+Q_2,0}.
\end{equation}

\section{Bulk realisation of conformal Carroll symmetries }\label{bulk}
In this section, we turn to the bulk realisation of conformal Carroll symmetries. We identify spacetimes whose isometry algebra is the type-D conformal extension of the Carroll algebra\footnote{The type-K conformal extension will be discussed in the conclusion.}, with $|z| \neq + \infty$. As mentioned in Section \ref{sec2}, following the results of \cite{Duval:2017els}, we look at four-dimensional plane wave spacetimes, which are known to possess $-$ at least $-$ the Carroll algebra without rotation as their isometry algebra. We start by providing a short introduction to plane waves before we study their isometries (see e.g. \cite{blau2011plane} for a review and an extensive list of references). As we will see, each of these plane wave spacetimes will be the vacuum solution of a phase space whose asymptotic symmetries are the infinite-dimensional $d=3$ conformal Carroll algebra for a given $z$. After defining these phase spaces, we comment on the special $z=1$ case, which corresponds to $\mathfrak{bms}_4$, and discuss the causal boundary of the plane waves at hand. Finally, we briefly extend our analysis to three-dimensional spacetimes.
 
\subsection{A lightning introduction to plane waves}
Plane wave spacetimes are a particular case of \textit{pp-wave} spacetimes, spacetimes that allow for a null, nowhere-vanishing and covariantly-constant vector field. They are most often described in two coordinate systems. The first one, called the \textit{Rosen} coordinate system, features two null coordinates $u$ and $v$ and a so-called \textit{wave profile} $C_{ij}(u)$:
\begin{equation}
	ds^2 = 2 du dv + C_{ij}(u) dy^i dy^j,
\end{equation}
where $y^i=\{x, y\}$ denotes space-like coordinates, transverse to the propagation of the wave. In Rosen coordinates, two different wave profiles might be related by a diffeomorphism. The second, called the \textit{Brinkmann} coordinate system, features one null coordinate $x^-$ and a wave profile $A_{ab}(x^+)$: 
\begin{equation}
	ds^2 = 2 dx^+ dx^- + A_{ab}(x^+) x^a x^b (dx^+)^2 + d\vec{x}^{\hspace{1pt}2}. 
\end{equation}
We refer to \textit{e.g.} Appendix A of \cite{Blau:2002js} for the transformation from Rosen to Brinkmann coordinates and vice-versa. This coordinate system has the advantage of unambiguously defining the spacetime through its wave profile. Plane wave spacetimes in four dimensions possess  \textit{at least} five isometries that correspond to the Carroll group in three dimensions without rotations \cite{Duval:2017els}. In Rosen coordinates, they act as 
\begin{align}
	u \to u, \quad y^i\to y^i+H_{ij}(u)b^{\,j}+c^i, \quad v \to v-b^{\,j}y_j-\frac{1}{2}b^{\,j}H_{ij}(u)b_j+f.
\end{align}
From the perspective of the Carroll group, $b^i$ parametrises two Carroll boosts and $c^i,\, f$, two spatial and a temporal translations, respectively. The matrix $H_{ij}(u)$ can be determined in terms of $C_{ij}(u)$ as 
\begin{equation*}
	H_{ij}(u)=\int (C_{ij}(u))^{-1} du.
\end{equation*} 
When the plane wave is conformally flat, which is equivalent to having
\begin{equation}
	A_{ab}(x^+) = A(x^+) \delta_{ab},
\end{equation}
rotations are restored. Depending on the wave profile $A_{ab}(x^+)$, an additional isometry with a component along $\partial_+$ can exist, potentially enhancing the number of symmetries to seven. As a special case, note that a vanishing Brinkmann wave profile corresponds to flat space in double null coordinates, for which the isometry group is exceptionally enhanced to ten generators.  Furthermore, all the (polynomial) curvature invariants of plane-wave spacetimes are zero \cite{blau2011plane}. \\

Plane wave spacetimes are not necessarily solutions of the vacuum Einstein equations. They are when the Brinkmann wave profile is traceless, which implies that no plane wave solution of Einstein gravity can have seven isometries. In what follows, we will pay special attention to the plane-wave metric
\begin{equation}
ds^2 = 2 du dv + e^u (dx^2 + dy^2) \label{z=0metric}
\end{equation}
and to the one-parameter family of geometries
\begin{equation}
	ds_{(k)}^2 = 2 du dv + u^k (dx^2 + dy^2), \qquad k\in \mathbb{R}. \label{Eq:kmetricRosen}
\end{equation}
In Brinkmann coordinates, these can be written as 
\begin{equation}
	ds^2= 2 dx^+ dx^- + \dfrac{(x^2 + y^2) }{4}(dx^+)^2 + dx^2 + dy^2 \label{Eq:z=0brinkmann}
\end{equation}
and
\begin{equation}
ds_{(k)}^2 =  2 dx^+ dx^- + \dfrac{k(k-2)}{4(x^+)^2}(x^2 +y^2) (dx^+)^2 + dx^2 + dy^2. \label{Eq:kprofileBrinkmann}
\end{equation}
The metric \eqref{Eq:z=0brinkmann} can be shown to be the Penrose limit \cite{Penrose}, \textit{i.e.} the geometry in the infinitesimal neighbourhood of null geodesics, of dS$_2 \times H^2$ \cite{Singh_2005}. Plane waves with Brinkmann wave profiles $A_{ab}(x^+) \sim (x^+)^{-2}$ are called \textit{scale-invariant} plane waves as they are invariant under the scale transformation
\begin{equation}
	(x^+, x^-, x^i) \to ( \lambda x^+, \lambda^{-1} x^-, x^i).
\end{equation} 
This particular power-law behaviour of the wave profile has been shown to be a feature of Penrose limits around singularities \cite{Blau_2004}, but also corresponds to the Penrose limit of FRW spacetimes or the near-horizon regions of fundamental strings and Dp-brane backgrounds \cite{Blau_2002, Fuji_2002}. In Rosen coordinates, the only non-vanishing component of the Ricci tensor is 
\begin{equation}
    R_{uu} = -\dfrac12 \qquad \text{ and } \qquad R^{(k)}_{uu}= -\dfrac{k(k-2)}{2u^2}, \label{Eq:Ricci}
\end{equation}
respectively. It is known that plane waves are exact backgrounds of string theory \cite{Amati:1988sa, Horowitz:1990sr}. In the string frame, the action
\begin{equation}
	\mathcal{S}= \int d^4x \, \sqrt{-g} \, e^{-2\phi}\left(R - \dfrac{1}{12} H^2 + 4 \nabla_\mu \phi \nabla^\mu \phi \right)
	\label{Eq:sigmamodel}
\end{equation} 
results in the following equations of motion: 
\begin{align}
	&R - \dfrac{1}{12} H^2 -4 \nabla_\mu \phi \nabla^\mu \phi + 4 \nabla_\mu \nabla^\mu\phi =0, \\
	&R_{\mu \nu}= \dfrac14 H_{\mu \nu}^2 - 2 \nabla_\mu \nabla_\nu \phi, \label{Eq:EOMString2}\\
	&\nabla_\mu H^{\mu \nu \rho} =2 \nabla_\mu \phi H^{\mu \nu \rho}.
\end{align}
Given Eqs. $\eqref{Eq:Ricci}$ and $\eqref{Eq:EOMString2}$, we see that the metric \eqref{z=0metric} cannot be supported by a constant dilaton. Likewise, for $k<0$ and $k>2$, a varying dilaton is required to support \eqref{Eq:kprofileBrinkmann}. For plane waves with $C_{ij}(u)=C(u)\delta_{ij}$, if we assume that the dilaton is an arbitrary function of $u$, then the equations of motion result in $H_{uxy}\equiv H_{uxy}(u)$ being the only non-zero component of the flux form. The only additional constraint provided by the equations of motion is the second-order differential equation
\begin{equation}
	R_{uu} - \dfrac{1}{2}(C(u))^{-2} H_{uxy}^2 + 2 \phi''(u)=0. 
\end{equation}
Hence, for a given plane wave, one can simply fix either the gauge field or the dilaton and determine one in terms of the other. Note that when $0<k<2$, the Ricci tensor is positive and can thus be compensated by a flux satisfying $H_{uu}^2= 4 R_{uu}$.

\subsection{Bulk realisation of $z=0,\, d=3$ conformal Carroll symmetry}
Let us first focus on the geometry described by
\begin{equation}
ds^2 = 2 du dv + e^u (dx^2 + dy^2) .
\end{equation}
This plane wave possesses a seven-dimensional isometry group: the full three-dimensional Carroll group and an additional symmetry with a component along $\partial_u$. One can promote the Euclidean plane spanned by $x$ and $y$ to the complex plane by introducing the complex coordinates 
\begin{equation}
	z= x+ iy, \qquad \bar{z}= x-iy.
\end{equation}
In these coordinates, the isometry generators are 
\begin{equation}
	\begin{aligned}
		L_0       &= -\partial_u + z \partial_z,                &\qquad \bar{L}_0       &= -\partial_u + \bar{z} \partial_{\bar{z}}, \\
		L_{-1}    &= -\partial_z,                                &\qquad \bar{L}_{-1}    &= -\partial_{\bar{z}}, \\
		M_{1,0}   &= z \partial_v + 2 e^{-u} \partial_{\bar{z}}, &\qquad M_{0,1}         &= \bar{z} \partial_v + 2 e^{-u} \partial_z, \\
		M_{0,0}   &= \partial_v.
	\end{aligned}
\end{equation}
Not only is the algebra spanned by these vector fields isomorphic to the minimal extension of the $d=3$ Carroll algebra preserving the Carrollian vacuum introduced in Section \ref{sec2}, i.e. \eqref{typeD3d}, but they reduce to the field theory generators when we let $u \to + \infty$, limit that we motivate further. Fixing this coordinate defines a codimension-one null hypersurface in the plane-wave spacetime. This hypersurface is degenerate, and its non-degenerate transverse spatial submetric $h_{ab}$, with $a,b=z, \bar{z}$, is flat. Endowed with the nowhere-vanishing null vector field $\partial_v$ and $h_{ab}$, this hypersurface defines a Carroll structure. These observations lead us to contemplate the possibility of a holographic duality between gravity in asymptotically plane-wave spacetimes and conformal Carrollian field theories defined by vacuum C, as introduced in Section \ref{Sec:CorrFct}, on this codimension-one hypersurface where $\partial_v$ defines the Carrollian time. In order to explore this prospect, we start by defining the $z=0$ plane-wave phase space with the gauge conditions
\begin{equation}
	g_{uu}= g_{vv}=0, \quad g_{uv}=1, \quad \partial_u \det \left(e^{-u} h_{ab} \right)=0.
\end{equation}
Along with them, the most general geometry takes the form 
\begin{equation}
	\hspace{-5 pt}ds^2 = 2 du dv + A_a(u,v,z,\bar{z}) du dx^a + B_a(u,v,z,\bar{z}) dv dx^a + h_{ab}(u,v,z,\bar{z}) dx^a dx^b, \label{Eq:GeneralGauge}
\end{equation}
where $x^a=(z, \bar{z})$ and both $A_a$ and $B_a$ are free functions of all coordinates. The spatial metric $h_{ab}$ is however subject to a determinant condition. The $z=0$ plane-wave phase space will be defined by the following restriction of \eqref{Eq:GeneralGauge}: 
\begin{equation}
	ds^2 = 2 du dv + A_a(z,\bar{z}) du dx^a + B_a(z,\bar{z}) dv dx^a + h_{ab}(u,z,\bar{z}) dx^a dx^b,
\end{equation}
with 
\begin{equation}
	h_{zz}=h_{zz}(z,\bar{z}), \quad 	h_{\bar{z}\bar{z}}=h_{\bar{z}\bar{z}}(z,\bar{z}), \quad h_{z \bar{z}}= \dfrac{e^u}{2} + h(z,\bar{z}). \label{z=0phasespace}
\end{equation}
The above class of metrics reduces to metric \eqref{z=0metric} for 
\begin{equation}
	A_a=B_a=h_{zz}=h_{\z\z}=h=0.
\end{equation}
This landscape of geometries contains more than just plane waves or pp-waves. We can identify the pp-wave sector of our phase space in the following manner: pp-waves admit a covariantly-constant, nowhere-vanishing, null Killing vector. Obviously, $\xi^\mu \partial_\mu =\partial_v$ is a nowhere-vanishing null Killing vector for \eqref{z=0phasespace}. The requirement that is covariantly constant is then equivalent to 
\begin{equation}
	\nabla_{\mu}\xi_{\nu}=\nabla_{\nu}\xi_{\mu},
\end{equation}
which leads to the condition
\begin{equation}
	 \label{pp}
	\partial_a g_{vb}= \partial_b g_{va}
	\quad \Leftrightarrow \quad g_{va}=\partial_a \Phi(z,\z),
\end{equation}
with $\Phi$ an arbitrary function of the complex coordinates. In what follows, we will not restrict ourselves to this sector. One can verify straightforwardly that this condition does not restrict the group of symmetries any further. We now look for the diffeomorphisms $\xi$ preserving the phase space \eqref{z=0phasespace}. Gauge conditions impose ${\mathcal{L}_\xi g_{uu}=\mathcal{L}_\xi g_{vv}=\mathcal{L}_\xi g_{uv}=0}$. The first two constraints result in 
\begin{equation}
    \partial_u \xi^v = \partial_v \xi^u=0,
\end{equation}
whereas the third can be rewritten as 
\begin{equation}
    \partial_u \xi^u + \partial_v \xi^v=0.
\end{equation}
Since $\xi^u\equiv \xi^u(u,z,\bar{z})$ and $\xi^v\equiv \xi^v(v,z,\bar{z})$, we find 
\begin{equation}
    \xi^u(u,z, \bar{z})= u \xi^u_{(1)}(z, \bar{z}) + \xi^u_{(0)}(z, \bar{z}), \quad \xi^v(v,z, \bar{z})= -v \xi^u_{(1)}(z, \bar{z}) + \xi^v_{(0)}(z, \bar{z}).
\end{equation}
We now move on to the other components of the metric.
We have
\begin{equation}
    \mathcal{L}_\xi g_{ua} = \xi^\alpha \partial_{\alpha} g_{u a} + \partial_u \xi^\alpha g_{\alpha a} + \partial_a \xi^\alpha g_{u \alpha}. 
\end{equation}
We require this be a function of the spatial coordinates $z$ and $\bar{z}$ only. The only $u-$ or $v-$dependent contributions to this equation that might appear stem from $\xi^u$, $\xi^v$ or $h_{z \bar{z}}$. However, any term involving $\xi^u$ will either have its undesired dependence cancelled by derivatives acting on it or will vanish due to the gauge conditions. The same goes for $\xi^v$, except for the term $\partial_a \xi^v g_{uv}$, which implies 
\begin{equation}
     \xi^u(u,z, \bar{z})= C u  + \xi^u_{(0)}(z, \bar{z}), \quad \xi^v(v,z, \bar{z})= -C v  + \xi^v_{(0)}(z, \bar{z}),
\end{equation}
with $C$ a constant. The remaining problematic contributions vanish if we set
\begin{equation}
    \xi^a\equiv \xi^a(z,\bar{z}).
\end{equation}
Requiring the variation of $g_{va}$ to also be a function of the spatial coordinates only does not constrain the diffeomorphism any further. Furthermore, since $\partial_u h_{z\bar{z}}(u,z,\bar{z})= e^u$, we also have 
\begin{equation}
\begin{aligned}
    \mathcal{L}_\xi h_{z \bar{z}} &= \xi^\alpha \partial_\alpha g_{z \bar{z}} + \partial_z \xi^\alpha g_{\alpha \bar{z}} + \partial_{\bar{z}} \xi^\alpha g_{z  \alpha} \\
    &= \left(-C u + \xi_{(0)}^u\right)  e^u  + \left(\partial_z \xi^z + \partial_{\bar{z}} \xi^{\bar{z}}\right)\left(e^u + h(z,\bar{z})\right)+ \dots 
    \end{aligned}
\end{equation}
Again, we ask this variation to be a function of the spatial coordinates only. Then, we find $C=0$ and
\begin{equation}
    \xi^u_{(0)} + (\partial_z \xi^z + \partial_{\bar{z}} \xi^{\bar{z}})=0.
\end{equation}
Finally, the variation of the diagonal elements of the spatial metric imply 
\begin{equation}
    \xi^z \equiv f^z(z), \quad \xi^{\bar{z}}\equiv f^{\bar{z}}(\bar{z}).
\end{equation}
All in all, with $\xi^v_{(0)} \equiv \alpha(z, \bar{z})$, we find that the diffeomorphism preserving the $z=0$ phase space can be written in terms of arbitrary functions: 
\begin{align} \label{asymptotic}
	\xi=-\big(\p_z f^z(z)+\p_{\bar{z}}f^{\bar{z}}(\z)\big)\p_u+\alpha(z,\z)\p_v+f^z(z)\p_z+f^{\z}(\z)\p_{\z}
\end{align}
where $\alpha(z,\z), f^z(z)$ and $f^{\z}(\z)$ respectively account for supertranslations and superrotations. We can expand the arbitrary functions in Laurent series: 
\begin{equation}
    f^z(z)= -\sum_n z^{n+1} f_n, \quad  f^{\bar{z}}(z)= -\sum_n z^{n+1} \bar{f}_n, \quad \alpha(z, \bar{z}) = \sum_{r,s} z^r \bar{z}^s \alpha_{r,s}. 
\end{equation}
This defines the generators
\begin{equation}
	 L_n = (n+1) z^n \partial_u - z^{n+1} \partial_z,  \quad 
	\bar{L}_n = (n+1) \bar{z}^{n} \partial_u - \bar{z}^{n+1} \partial_{\bar{z}}, \quad 
	M_{r s}= z^r \bar{z}^s \partial_v,
\end{equation}
whose commutation relations are those of the $d=3$, $z=0$ conformal Carroll algebra, which is what we wanted to achieve.

\subsection{Bulk realisation of $0<z<\infty, \, d=3$ conformal Carroll symmetries}
As mentioned in Section \ref{sec3}, the $0<z=2/k<+\infty$ conformal Carroll algebra is generated by the following vector fields: 
\begin{equation}
    L_n= -z^{n+1} \partial_z - \dfrac{2(n+1)}{k} z^n t \partial_t, \quad  \bar{L}_n= -\bar{z}^{n+1} \partial_{\bar{z}} - \dfrac{2(n+1)}{k} \bar{z}^n t \partial_t, \quad M_{r,s}= z^r \bar{z}^s \partial_t.
\end{equation}
These generators obey the following non-trivial commutation relations:
\begin{equation}
\begin{aligned}
    [L_n, L_m]= (n-m) L_{m+n}, \quad [L_n, M_{r,s}] = \left(\dfrac{n+1}{k} -r \right) M_{r+n,s},\\
     [\bar{L}_n, \bar{L}_m]= (n-m) \bar{L}_{m+n}, \quad [\bar{L}_n, M_{r,s}] = \left(\dfrac{n+1}{k} -s \right) M_{r,s+n}.
     \end{aligned}
\end{equation}
Again, we find a seven-dimensional subalgebra spanned by $\{L_0,\bar{L}_0,L_{-1},\bar{L}_{-1},M_{0,0},M_{1,0},M_{0,1}\}$ containing the three-dimensional Carroll algebra and a $z\neq 0$ dilation. Note that the $z\neq 0$ analogue of the subset of symmetries preserving vacuum W does not respect algebraic closure anymore. \\

  We now turn to the one-parameter family of metrics \eqref{Eq:kmetricRosen}. The $k=0$ and $k=2$ cases correspond to flat space, which we will cover separately. For $k\neq1$, the vector fields generating the isometry algebra of \eqref{Eq:kmetricRosen} are
 \begin{equation}
	\begin{aligned}
		L_0       &= \dfrac{u}{k} \partial_u - \dfrac{v}{k} \partial_v - z \partial_z,            &\qquad \bar{L}_0    &= \dfrac{u}{k} \partial_u - \dfrac{v}{k} \partial_v - \bar{z} \partial_{\bar{z}}, \\
		L_{-1}    &= -\partial_z,                                                                &\qquad \bar{L}_{-1} &= -\partial_{\bar{z}}, \\
		M_{1,0}   &= z \partial_v + \dfrac{2}{k-1} u^{1-k} \partial_{\bar{z}},                   &\qquad M_{0,1}      &= \bar{z} \partial_v + \dfrac{2}{k-1} u^{1-k} \partial_z, \\
		M_{0,0}   &= \partial_v.
	\end{aligned}
\end{equation}
Analogously to the $z=0$ case, this is isomorphic to the seven-dimensional subalgebra identified hereabove. Again, this shows that on a carefully chosen codimension-one hypersurface, the isometry generators reduce exactly to the generators of this seven-dimensional subalgebra with $v$ being the Carrollian time. The choice of this hypersurface is dictated by the value of $k$: for $k>1$, $u \to +\infty$ will cancel unwanted contributions to the isometry generators whereas for $k<1$, we send $u \to 0$. The $k=1$ case is special, as $M_{0,1}$ and $M_{1,0}$ contain a $\log u$ term which replaces the $u^{k-1}/(k-1)$ factor that appears for general $k\neq 1$. For this particular value of $k$ only, we set $u=1$ as the relevant hypersurface. We now proceed in the same manner as in the previous section and pick a gauge (this is now valid for any $k$):
\begin{equation}
    g_{uu}=g_{vv}=0, \quad g_{uv}=1, \quad \partial_u \det\left(u^{-k}h_{ab} \right)=0,
\end{equation}
with $h_{ab}$ the transverse metric. Note that for $k=2$, that is $z=1$, the determinant condition is equivalent to that of the Bondi gauge for $u \sim r$. Again, in this gauge, the most general metric is written as \eqref{Eq:GeneralGauge}. We define the $z\neq 0$ phase space as the following restriction on the arbitrary functions appearing therein
\begin{equation}
\begin{aligned}
    A_a \equiv A_a(v,z, \bar{z}), \quad &B_a\equiv u B_a(z,\bar{z}), \quad h_{zz}\equiv u h_{zz}(v,z, \bar{z}), \quad h_{\bar{z}\bar{z}}\equiv u h_{\bar{z}\bar{z}}(v,z, \bar{z}),\\
    & \qquad  \qquad h_{z \bar{z}} \equiv \dfrac{u^k}{2} + u h(v,z,\bar{z}),
    \end{aligned}
\end{equation}
and look for the symmetries preserving this phase space. Again, note that the metric \eqref{Eq:kmetricRosen} corresponds to setting all arbitrary functions to zero. As this computation is almost identical to the one carried out for $z=0$, we do not expose the details of our calculations. A notable difference between the $z=0$ case and this result is that linear contributions in $u$ and $v$ do not vanish. Furthermore, since $\partial_u u^k= k u^{k-1}$, the diffeomorphism will explicitly involve $k$. We find this phase space to be preserved by
\begin{equation}
   \xi = -\dfrac{1}{k} \left(\partial_z f^z(z) + \partial_{\bar{z}} f^{\bar{z}}(\bar{z}) \right) \left(u \partial_u - v \partial_v \right) + \alpha(z, \bar{z}) \partial_v + f^z(z) \partial_z + f^{\bar{z}}(\bar{z}) \partial_{\bar{z}}.
\end{equation}
Expanding the arbitrary functions in $\xi$ in modes, we find the following generators 
\begin{equation}
	\begin{aligned}
		L_n&= \dfrac{n+1}{k} z^n \left( u\partial_u - v\partial_v \right)  - z^{n+1} \partial_z, \\
		\bar{L}_n&= \dfrac{n+1}{k} \bar{z}^n \left( u\partial_u - v\partial_v \right)  - \bar{z}^{n+1} \partial_{\bar{z}}, \\
		M_{r,s}&= z^r \bar{z}^s \partial_v. 
	\end{aligned}
\end{equation}
One can verify that these generators obey the commutation relations of the \linebreak ${d=3, 0 < z=2/k < + \infty}$ conformal Carroll algebra.

\subsection{The special case of flat space}
It is well-known that the $z=1$ conformal Carroll algebra is isomorphic to the BMS algebra in one higher dimension \cite{Duval_2014}. As previously mentioned, the $k=0$ and $k=2$ cases of \eqref{Eq:kmetricRosen} correspond to flat space. Let us work with $k=2$, as $k=0$ yields a trivial wave profile. The isometry algebra is now ten-dimensional and contains the following additional vector fields: 
\begin{equation}
	\begin{aligned}
		L_1 &= u z \partial_u - v z \partial_v - z^2 \partial_z - 2 \dfrac{v}{u} \partial_{\bar{z}}, \qquad 
		\bar{L}_{1} = u \bar{z} \partial_u - v \bar{z} \partial_v - \bar{z}^2 \partial_{\bar{z}} - 2 \dfrac{v}{u} \partial_{z},\\
		& \qquad \qquad \qquad \qquad  M_{1,1}= -2 \partial_u + z \bar{z} \partial_v + \dfrac{2 z}{u} \partial_z + \dfrac{2 \bar{z}}{u} \partial_{\bar{z}},
	\end{aligned}
\end{equation}
which also reduce to the exact corresponding field theory symmetry generators $\{L_1, \bar{L}_1, M_{1,1}\}$ as we take $u \to + \infty$. In this particular case, we can justify the choice of the boundary value for $u$ by transforming directly to Bondi coordinates. First, we go from the $k=2$ coordinate system to the $k=0$ coordinate system using
	\begin{equation}
		x= \dfrac{x'}{u'}, \quad y= \dfrac{y'}{u'}, \quad u=u',\quad  v= v' + \dfrac{1}{2 u'}\left((x')^2 + (y')^2\right),
	\end{equation}
	which brings the metric to 
	\begin{equation}
		ds^2 = 2 du' dv' + (dx')^2 + (dy')^2.
	\end{equation}
	Then, one goes back to the standard Minkowski metric in Cartesian coordinates by redefining
	\begin{equation}
		 u'= \dfrac{1}{\sqrt{2}}(t-z), \quad v' = -\dfrac{1}{\sqrt{2}}(t+z).
	\end{equation}
   All in all, one goes from the original coordinate system $\{u,v,z,\bar{z}\}$ of the $k=2$ plane wave to the retarded Bondi coordinates $\{\tilde{u},r,\theta, \varphi\}$  for flat space with 
    \begin{equation}
  	u = \dfrac{1}{\sqrt{2}} F(r,\tilde{u}, \theta), \quad v = -\dfrac{1}{\sqrt{2}} \dfrac{\tilde{u}(2 r + \tilde{u})}{F(r,\tilde{u}, \theta)}, \quad z= \sqrt{2} \dfrac{r e^{i \varphi}\sin \theta}{F(r,\tilde{u},\theta)},
  \end{equation}
  where $F(r,\tilde{u},\theta)$ is defined as $F(r,\tilde{u}, \theta)= r(1-\cos \theta) + \tilde{u}$. Then, as one reaches $\mathscr{I}^+$, that is $r\to + \infty$, we find that $\tilde{u}\to \infty$ and all the other coordinates are finite, which justifies our choice of asymptotic region for the $z=1$, or $k=2$ case. For $z=1$, our definition of the phase leads us to the following boundary conditions: 
  \begin{equation}
    g_{\mu \nu}^{\text{B.C.}}=  \begin{pmatrix}
      0 & 1 & A_z(v,z,\z) + \mathcal{O}(u^{-1}) & A_\z(v,z,\z) + \mathcal{O}(u^{-1}) \\
      \times & 0 & u \,B_z(z, \z) + \mathcal{O}(u^0) & u\, B_\z(z, \z) + \mathcal{O}(u^0) \\
      \times & \times & u \,h_{zz}(v,z, \z) + \mathcal{O}(u^0) & \frac{u^2}{2} + u \,h(v,z,\z) + \mathcal{O}(u^0) \\
      \times & \times & \times & u\, h_{\z \z}(v,z, \z) + \mathcal{O}(u^0)
  \end{pmatrix}.
  \end{equation}

  \subsection{Conformal flatness and the conundrum of plane wave boundaries}
  All the plane wave metrics presented in this section are conformally flat, as their wave profiles are proportional to the identity. Hence, it should be possible to write them, at least locally, as 
  \begin{equation}
  ds^2= \Omega^2(x) \eta_{\mu \nu},
  \end{equation}
  where $\Omega^2(x)$ is a strictly positive conformal factor. Furthermore, it was shown that the causal boundary of a plane wave with constant wave profile is a one-dimensional null line parametrised by $x^+$ in Brinkmann coordinates, first for the maximally-supersymmetric Nappi-Witten plane wave in \cite{berenstein2002lightconestringfieldtheory} and then for more general wave profiles in \cite{Hubeny_2002, Marolf_2002}. However, as mentioned in \cite{Marolf_2002}, this conclusion only applies to non-positive-definite wave profiles $A_{ab}(x^+)$. Let us review an argument from \cite{Marolf_2002} and consider once again the metric
  \begin{equation}
      ds^2= 2 du dv + e^u (dx^2 + dy^2).
  \end{equation}
As shown before, in Brinkmann coordinates, this is simply (note that the $1/4$ factor multiplying $(dx^+)^2$ was reabsorbed for convenience):
\begin{equation}
    ds^2 = 2 dx^+ dx^- + (x^2 + y^2) (dx^+)^2 + dx^2 + dy^2. 
\end{equation}
The change of coordinates 
\begin{equation}
    \begin{aligned}
        \tilde{v}&= -x^- - (x^2 + y^2) \tanh{x^+}, \\
        \tilde{u}&= -\tanh{x^+}, \\
        \tilde{x}&= x/ \cosh{x^+}, \\
        \tilde{y}&= y/ \cosh{x^+}
    \end{aligned}
\end{equation}
results in the metric 
\begin{equation}
    ds^2= \dfrac{1}{1-\tilde{u}^2} \left(2 d\tilde{u} d\tilde{v} + d\tilde{x}^2 + d\tilde{y}^2 \right), 
\end{equation}
which is conformally equivalent to a slice of Minkowski space bounded by two null planes at $\tilde{u}= \pm 1$, that is for $x^+ = \pm \infty$ or equivalently, for $u = \pm \infty$ in the original Rosen coordinates. Hence, this is direct justification of the $u \to \infty$ limit we have taken to project the isometry generators of the metric \eqref{z=0metric} to the three-dimensional $z=0$ field theory generators. Furthermore, the codimension-one (degenerate) hypersurface described by fixing $u \to \infty$ is actually the causal boundary of the pp-wave spacetime, which comforts us in the idea of a holographic duality in the traditional sense.\\

  Let us now move on to the other pp-wave spacetimes at hand, namely \eqref{Eq:kmetricRosen}, for which this discussion is more subtle. As mentioned before, spacetimes with $A_{ab}(x^+) \sim (x^+)^{-2}$ can be viewed as Penrose limits of singular spacetimes. Whatever the original nature of the singularity, the Penrose limit procedure yields a null singularity in the resulting plane wave spacetime. The geodesic incompleteness of the parent spacetime is inherited by the plane wave and for this reason, $x^+$, or $u$, cannot be extended to the whole real axis. Before discussing the causal boundary, let us remind the reader of the relevant family of plane wave metrics in Brinkmann coordinates: 
\begin{equation}
    ds_{(k)}^2 = 2 dx^+ dx^- + \dfrac{k(k-2)}{4 (x^+)^2} (x^2 + y^2) (dx^+)^2 + dx^2 + dy^2.
\end{equation}
Notice that this family of metrics possesses a reflection axis at $k=1$. Then, with $\alpha>0$, we can straightforwardly conclude that the $z=2/(1+\alpha)$ and $z=2/(1-\alpha)$ type-D subalgebras are isomorphic, as they are the isometry algebra of the same geometry. In Rosen coordinates, one goes from 
\begin{equation}
    ds^2 = 2 du dv + u^{1+ \alpha} (dx^2 + dy^2)
\end{equation}
to 
\begin{equation}
    ds^2 = 2 du dv' + u^{1- \alpha} (dx'^2 + dy'^2)
\end{equation}
by performing the change of coordinates:
\begin{equation}
   u'= u, \quad  v'= v- \dfrac{\alpha}{2} u^{\alpha} (x^2 + y^2), \quad x' = u^\alpha x, \quad y'= u^\alpha y.
\end{equation}
From the point of view of the $z\neq0$ conformal Carroll algebra alone, one can verify that if some vector fields $\{L_n, \bar{L}_n, M_{r,s}\}$ satisfy the $z=2/(1+\alpha)$ conformal Carroll algebra, then the generators
\begin{equation}
  \hspace{20 pt}  \begin{aligned}
       & L_0'= \dfrac{1}{1-\alpha}(\alpha L_0 + \bar{L}_0), \qquad   \bar{L}_0'= \dfrac{1}{1-\alpha}(\alpha \bar{L}_0 + L_0), \\
      & \hspace{-55 pt} L_{-1}' = M_{1,0}, \quad  \bar{L}_{-1}' = M_{0,1},\quad  M_{1,0}'= L_{-1}, \quad M_{0,1}'= \bar{L}_{-1}, \quad M_{0,0}'=M_{0,0}
    \end{aligned}
\end{equation}
satisfy the $z=2/(1-\alpha)$ type-D Carroll algebra. That being said, let us work with \eqref{Eq:kmetricRosen} and exploit this reflectivity property to halve our efforts:
\begin{equation}
    ds_{(k)}^2= 2 du dv + u^k(dx^2 + dy^2), \end{equation}
where $u \in (0, + \infty)$, and take $k>1$. We define the new (null) coordinate
    \begin{equation}
        u= \left((k-1)\tilde{u}\right)^{\frac{1}{1-k}}. 
    \end{equation}
    Since $k>1$, this coordinate change sends $u=0$ to $\tilde{u}= + \infty$ and conversely, $u \to + \infty$ is mapped to $\tilde{u}=0$. Upon redefining $v \to -v$, we find that the metric can be seen as conformally related to a $\tilde{u}>0$ strip of Minkowski space:
    \begin{equation}
        ds^2_{k>1} = (k-1)^{\frac{k}{1-k}} \tilde{u}^{\frac{k}{1-k}} (2 d\tilde{u} dv + dx^2 + dy^2).
    \end{equation}
    In order to analyse the causal boundary of that spacetime in the usual compactification scheme, we wish to recast the Minkowski metric as a conformal factor times the Einstein static universe. We partly use the notations of \cite{Papadopoulos_2003}. With 
    \begin{equation}
        \tilde{u}= \dfrac{ \sin \phi + \cos \theta \sin \psi}{\cos \phi + \cos \psi}, \label{Eq:uEinsteinStatic}
    \end{equation}
    where $0\leq \psi \leq \pi$, $0\leq \theta \leq \pi$, we find 
    \begin{equation}
        ds^2_{k>1}= (k-1)^{\frac{k}{1-k}} \dfrac{(\sin \phi + \cos \theta \sin \psi)^{\frac{k}{1-k}}}{(\cos \phi + \cos \psi)^{\frac{k-2}{k-1}}} \left( -d\phi^2 + d\psi^2 + \sin^2 \psi (d\theta^2 + \sin^2 \theta^2 d\alpha^2)\right). \label{Eq:ConfFactor}
    \end{equation}
Let us study the conformal factor. We distinguish two cases: 
\begin{itemize}
    \item $1 < k \leq 2 $: the denominator of \eqref{Eq:ConfFactor} carries a negative (or zero) power and thus will not contribute to making the conformal factor diverge. However, since the numerator carries a negative exponent as well, we conclude that the conformal boundary is defined by 
    \begin{equation}
        \sin \phi + \cos \theta \sin \psi=0,
    \end{equation}
    which coincides with a three-dimensional subspace of the Einstein static universe (except for special points that are not relevant to our analysis). Given Eq. \eqref{Eq:uEinsteinStatic}, this corresponds to $\tilde{u}=0$, or $u \to +\infty$. 
    \item $k>2$: the exponent in the denominator of Eq. \eqref{Eq:ConfFactor} is positive and a second boundary appears. This new boundary is defined by 
    \begin{equation}
        \cos \phi + \cos \psi =0,
    \end{equation}
    that is $\tilde{u}\to +\infty$, or $u=0$. 
\end{itemize}
Since we have already established that the family of plane waves that we consider is reflective around $k=1$, what we find is consistent with the analysis carried out in \cite{Papadopoulos_2003} for $0<k<1$. Based on the projection of the isometries of \eqref{Eq:kmetricRosen} on the symmetry generators of a putative dual field theory, we argued that the relevant hypersurface for holography was $u \to + \infty$ for $k>1$ and $u=0$ for $k<1$. In the latter case, we have just shown that for $0<k<1$, this hypersurface is not the conformal boundary of the associated plane wave spacetime. Let us now comment on the peculiar $k=1$ case, whose  metric is 
\begin{equation}
    ds^2_{k=1} = 2 du dv + u (dx^2 + dy^2). 
\end{equation}
By defining $u= e^{\tilde{u}}$, that is $\tilde{u}= \log u \in \mathbb{R}$, we find 
\begin{equation}
\begin{aligned}
     ds^2_{k=1} &= e^{\tilde{u}} (2 d\tilde{u} dv + dx^2 + dy^2)  \\
     &= \dfrac{\exp{\left(\frac{ \sin \phi + \cos \theta \sin \psi}{\cos \phi + \cos \psi}\right)}}{(\cos \phi + \cos \psi)^2}\left( -d\phi^2 + d\psi^2 + \sin^2 \psi (d\theta^2 + \sin^2 \theta^2 d\alpha^2)\right). 
\end{aligned}
\end{equation}
The conformal factor obviously diverges for $\cos \phi + \cos \psi=0$, meaning $\tilde{u}\to \infty$, \textit{i.e.} $u=0$ or $u\to + \infty$. This is also different from the hypersurface we chose purely on the basis of projecting the isometries of the $k=1$ plane wave onto the field theory generators.

\subsection{Comment on bulk Carrollian symmetries in $d=2$}
We briefly discuss the analogue of the above results of this section in one dimension less, $d=2$, which almost trivially amounts to get rid of the $y-$direction.

The metrics we consider are 
\begin{equation}
    ds^2= 2 du dv + e^u dx^2
    \label{d=3z=0}
\end{equation}
and 
\begin{equation}
    ds^2_{(k)}= 2 du dv + u^k dx^2. \label{d=3zgeneral}
\end{equation}
The metric \eqref{d=3z=0} corresponds to the Penrose limit of self-dual warped AdS$_3$ \cite{Fransen:2023eqj}. Its isometries are 
\begin{equation}
    \begin{aligned}
        L_0= 2\partial_u - x \partial_x, \qquad L_{-1}= - \partial_x, \qquad M_0= \partial_v, \qquad M_1= x \partial_v + e^{-u} \partial_x.
    \end{aligned}
\end{equation}
These four vector fields realise the $d=2$, $z=0$ type-D Carroll algebra. Likewise, for \eqref{d=3zgeneral}, the isometries are 
\begin{equation}
   \hspace{-3 pt} \begin{aligned}
        L_0= \dfrac{2}{k} (u\partial_u- v \partial_v) - x \partial_x, \quad L_{-1}= - \partial_x, \quad M_0= \partial_v, \quad M_1= x \partial_v  - \dfrac{u^{1-k}}{1-k} \partial_x,
    \end{aligned}
\end{equation}
which realise the $d=2$, $z=2/k$ type-D Carroll algebra. For the $z=0$ case, as was done in Section \ref{bulk}, we can work in the following gauge:
\begin{equation}
    g_{uu}=g_{vv}=0, \quad g_{uv}=1.
\end{equation}
The most general metric thus becomes 
\begin{equation}
ds^2 = 2 du dv +  A(u,v,x) du dx + B(u,v,x) dv dx + h_{xx}(u,v,x) dx^2.
\end{equation}
We define the $z=0$ phase space by specifying the  metric elements 
\begin{equation}
\begin{aligned}
    A \equiv A(x), \quad &B\equiv B(x), \quad h_{xx}\equiv e^u + h(x). \label{z=0d=2phase}
    \end{aligned}
\end{equation}
Then, we can show that the symmetries preserving the phase space \eqref{z=0d=2phase} are determined by two arbitrary functions
\begin{equation}
    \xi = -2 \partial_x f(x) \partial_u + \alpha(x) \partial_v + f(x) \partial_x.
\end{equation}
Upon mode expansion, we find the following three-dimensional bulk uplift of the $d=2$, $z=0$ conformal Carroll algebra: 
\begin{equation}
    L_n = 2(n+1) x^n \partial_u - x^{n+1} \partial_x, \qquad M_{r} = x^r \partial_v.
\end{equation}
Similarly, we define the $z=2/k$ phase space by specifying
\begin{equation}
\begin{aligned}
    A \equiv A(v,x), \quad &B\equiv u B(x), \quad h_{xx}\equiv u^k + u h(x).
    \end{aligned}
\end{equation}
This phase space is preserved by the following vector field:
\begin{equation}
   \xi = -\dfrac{2}{k} \partial_x f(x) + \left(u \partial_u - v \partial_v \right) + \alpha(x) \partial_v + f(x) \partial_x.
\end{equation}
When expanded in modes, we find 
\begin{equation}
    L_n = \dfrac{2(n+1)}{k} x^n (u \partial_u- v \partial_v) - x^{n+1} \partial_x, \qquad M_{r} = x^r \partial_v,
\end{equation}
which spans the $d=2$, $z=2/k$ conformal Carroll algebra and uplift its generators to the three-dimensional bulk.

\section{Conclusion and discussions}\label{conclusion}

In this article, we investigated the possibility of a holographic duality between certain plane wave spacetimes and a class of Carrollian conformal field theories. Since the Carrollian conformal algebra is isomorphic to the BMS algebra in one higher dimension, these theories have been considered as potential duals of asymptotically flat spacetimes. However, the independent scaling of space and time directions allows for a family of conformal extensions characterised by an exponent $z$, which we explored systematically. We analysed the structure of these general-$z$ Carrollian CFTs, with a particular focus on the $z=0$ case, and constructed bulk metrics in one higher dimension that realise the corresponding symmetries. In this context, we demonstrated that the four-dimensional plane wave metric \eqref{NWlike} exhibits a global subgroup spanned by seven generators of the $z=0$ algebra. By introducing a codimension-one null hypersurface as the boundary, we showed that the extrapolated bulk generators induce a conformal Carrollian structure on the intrinsic boundary geometry. To investigate if there is a class of spacetimes that  realises the infinite-dimensional extension of this finite subalgebra, we fixed a double-null gauge and defined a class of geometries that asymptotically approach the plane wave. The residual transformations preserving both the gauge and boundary conditions were shown to correspond to the infinitely-extended $z=0$ conformal Carroll symmetries. Finally, we outlined how this construction generalises to arbitrary values of $z$, providing a framework for further investigations into the holographic implications of Carrollian symmetries. Note that the approach we have taken is conceptually very different from the BMN limit of the AdS/CFT correspondence \cite{Berenstein:2002jq, berenstein2002lightconestringfieldtheory, Sadri:2003pr}, where one takes the Penrose limit of AdS$_5 \times S^5$ (the resulting metric being the ten-dimensional version of the Nappi-Witten metric \cite{Nappi:1993ie}) in order to reach a tractable regime of the superstring theory and identify the corresponding sector in the conformal field theory, whereas we have tried to set the foundations for a holographic duality involving plane waves as spacetimes of their own right, with a dual field theory living on their causal boundary. \\

\noindent There are several directions in which this work could be extended, which we list below:
\begin{itemize}
    \item We have identified spacetimes with seven isometries corresponding to the type-D conformal Carroll algebra. Another seven-dimensional conformal Carroll algebra of interest is the type-K conformal Carroll algebra mentioned in Section \ref{sec2}. During the preparation of this work, we have realised that the generators of this algebra appeared as the projection of the isometry generators of the Nappi-Witten metric \cite{Nappi:1993ie} 
    \begin{equation}
    \begin{aligned}
        ds^2&= 2 dx^+ dx^- - \dfrac{x^2+y^2}{4} (dx^+)^2 + dx^2 + dy^2, \label{Eq:NappiWitten} \\
        &= 2 du dv + \sin^2 u (dx^2 + dy^2)
    \end{aligned}
    \end{equation}
on well-chosen fixed-$x^+$ hypersurfaces. The causal boundary of this plane wave has been well-known for a long time, see for instance \cite{Marolf_2002, Hubeny_2002}, and is a one-dimensional null line parametrised by $x^+$. In that sense, these hypersurfaces are orthogonal to the causal boundary of the plane wave. We also mention that the Nappi-Witten metric \eqref{Eq:NappiWitten} can be seen as a double analytic continuation of the $z=0$ metric \eqref{z=0metric}.
\item The establishment of a holographic correspondence also relies on the matching of bulk correlators with those of the dual field theory. There have been several past efforts to match bulk correlators with non-relativistic field theory correlators by implementing the standard AdS/CFT procedure that identifies a normalisable and a non-normalisable mode \cite{BalaMcGreevy,Son:2008ye,Kachru:2008yh,Dong:2012se}. Because of the defining property of pp-waves, namely the existence of a covariantly-constant null Killing vector field, this same procedure cannot be implemented straightforwardly, as the scalar wave equation does not feature second-order derivatives in $u$ of Rosen coordinates (or equivalently, $x^+$ in Brinkmann coordinates). The question of whether coordinate systems where such modes can be identified and in which the boundary can be defined unambiguously has yet to be answered.

\item In Section \ref{bulk}, we presented families of metrics representing candidates for phase spaces realising the anisotropic conformal Carroll symmetries in two and three dimensions. It would be interesting to investigate whether these phase spaces lead to well-defined charge algebras and if potential central extensions appear, in the spirit of the study of bulk realizations of non-relativistic CFTs with Schr\"odinger(-Virasoro) symmetries \cite{Compere:2009qm}.

\item In this work, we identified candidates for bulk duals to anisotropic conformal Carroll field theories. We want to highlight that other geometries/boundary conditions with related properties have appeared in the literature. In three bulk dimensions, the so-called \textit{Warped Flat spacetimes}, solutions to Topologically Massive Gravity (TMG) \cite{Deser:1981wh,Deser:1982vy} have a vacuum with the four Killing vectors \eqref{VacuumC3D}. Their asymptotic symmetries were shown to form the $\mathfrak{ccar}_2^{(0)}$ algebra \cite{Detournay:2019xgl} in the first order formalism of TMG. In the framework of Rindler holography, boundary conditions reproducing the $\mathfrak{ccar}_2^{(0)}$ algebra were obtained in \cite{Donnay:2015abr, Afshar:2016wfy} (see Section 4.2 of \cite{Afshar:2016wfy} for a discussion of the global symmetries). In four bulk dimensions, boundary conditions with $\mathfrak{ccar}_3^{(0)}$
symmetries were introduced in \cite{Donnay:2015abr}, then generalized to $\mathfrak{ccar}_3^{(z)}$ (see \textit{e.g.} (A.8) of \cite{Grumiller_2020}) in the context of near-horizon symmetries, but without identifying potential vacuum metrics.
It would be interesting to investigate the connections between these geometries and various boundary conditions.

\item In two dimensions, we have discussed the distinction between WCFTs and $\mathfrak{ccar}_2^{(0)}$ field theories. In the bulk, this manifests itself in distinct bulk duals: WAdS$_3$ spaces in the former case \cite{Detournay:2012pc} and the backgrounds discussed in this paper and in the point above in the latter\footnote{Note that Warped flat spaces can be obtained as a contraction from Warped AdS$_3$ spaces, while the three-dimensional plane wave \eqref{d=3z=0} can be obtained as a Penrose limit from it.}. A natural question arising is then the following: what is, in four bulk dimensions, the bulk dual to Vacuum W \eqref{VacWC}? A natural guess could be AdS$_3 \times S^1$.

\end{itemize}

\section{Acknowledgements}

We thank the participants and organisers of the BITS Pilani Goa meeting ``Holography, Strings and other fun things'' for a stimulating atmosphere, as well as Arjun Bagchi, Aritra Banerjee, Rudranil Basu, Daniel Grumiller, Kevin Nguyen, Marc Geiller and Amitabh Virmani. We also thank M.M. Sheikh-Jabbari for discussions and comments on the manuscript. SD is a Senior Research Associate of the Fonds de la Recherche Scientifique F.R.S.-FNRS (Belgium).
SD acknowledges support of the Fonds de la Recherche Scientifique F.R.S.-FNRS (Belgium)
through the projects PDR/OL C62/5 “Black hole horizons: away from conformality” (2022-2025) and CDR n°40028632 (2025-2026). DF benefits from a FRIA fellowship granted by the F.R.S-FNRS. ED is a Research Fellow
of the Fonds de la Recherche Scientifique F.R.S.-FNRS (Belgium). This work is supported by the F.R.S.-FNRS (Belgium) through convention IISN 4.4514.08 and benefited from the support of the Solvay Family. 

\appendix
\section{Subgroups of $z$-BMS$_4$}
\label{appendix}
 From the $z=0$ case, we already have a hint that these algebras have two different classes of finite dimensional subalgebras. In this appendix, we refine the statement to hold for general $z$. The $z$-BMS$_4$ algebra is given by \cite{Safari_2019}
\begin{align}\label{z-bms}
	[L_n, L_m] &= (n - m) L_{n+m}, \quad &[\bar{L}_n, \bar{L}_m] &= (n - m) \bar{L}_{n+m}, \\
	[L_n, M_{r,s}] &= \left(z \frac{n+1}{2} - r\right) M_{r+n, s}, \quad 
	&[\bar{L}_n, M_{r,s}] &= \left(z \frac{n+1}{2} - s\right) M_{r, s+n}, \nonumber\\
	[M_{r,s}, M_{p,q}] &= 0. \nonumber
\end{align}
These $z$-BMS$_4$ algebras have a globally well-defined $SL(2,\mathbb{C})$ subgroup, spanned by the six generators $L_{0},L_{\pm1}$ and $\bar{L}_{0},\bar{L}_{\pm 1}$. We observe that the action of $L_{-1},\bar{L}_{-1}$ lowers the supertranslation indices, while $L_1$ and $\bar{L}_1$ raise them. The action of $L_0$ leaves the indices unchanged, \textit{i.e.}
\begin{align}
	[L_{-1},M_{r,s}]=-rM_{r-1,s}, \quad [L_0,M_{r,s}]=\left(\frac{z}{2}-r\right)M_{r,s}, \quad [L_1,M_{r,s}]=(z-r)M_{r+1,s}.
\end{align}  
This lowering and raising action truncates at $r=0$ and $r=z$, thus leading to a subgroup that contains the supertranslations $\{M_{0,0},M_{0,1},\dots,M_{z,z}\}$, along with $\{L_{0},L_{\pm1},\bar{L}_{0},L_{\pm 1}\}$. This subgroup is (6+$(z+1)^2$)--dimensional, which we denote as \textit{Type I}. For $z=1$, this subgroup is ten-dimensional and isomorphic to Poincaré algebra in four dimensions. \\

However, the algebra also admits infinitely many other finite-dimensional subgroups. Let us call them \textit{Type II} altogether. Out of the six $SL(2,\mathbb{C})$ generators, this class of subgroups only has $\{L_0,L_{-1},\bar{L}_0,\bar{L}_{-1}\}$. In the absence of the raising generators $L_1$ and $\bar{L}_1$, the algebra would close even with supertranslation generators up to an arbitrary positive index, as long as all the lower ones are also included. To be precise,
\begin{align}
	\{L_0,L_{-1},\bar{L}_0,\bar{L}_{-1},M_{0,0},M_{0,1},M_{1,0},\dots ,M_{p,q} \,| \, \forall p,q \in  \mathbb{Z}_{\geq 0}\}
\end{align}
forms a subgroup that is $(4+(p+1)(q+1))$--dimensional. This fact holds irrespective of the value of $z$.

\bibliographystyle{JHEP}
\bibliography{flat}

\end{document}